\documentclass[useAMS,usenatbib]{mn2e}
\usepackage{natbib}
\usepackage{color,graphicx} 
\usepackage{amsmath}        
\usepackage{amsfonts}       
\usepackage{amssymb}        
\usepackage{wasysym}        
\usepackage{marvosym}       
\usepackage{gensymb}        
\usepackage{framed}         

\long\def\symbolfootnote[#1]#2{\begingroup\def\thefootnote{\fnsymbol{footnote}}\footnote[#1]{#2}\endgroup}

\author[Keane et al.]{E.F.~Keane$^{1}$, M.~Kramer$^{1,2}$,
  A.G.~Lyne$^{1}$, B.W.~Stappers$^{1}$ \& M.A.~McLaughlin$^{3,4}$
  \\ $^{1}$ University of Manchester, Jodrell Bank Centre for
  Astrophysics, School of Physics \& Astronomy, Manchester M13 9PL,
  UK. \\ $^{2}$ Max Planck Insitut f\"{u}r Radioastronomie, Auf dem
  H\"{u}gel 69, 53121 Bonn, Germany. \\ $^{3}$ Department of Physics,
  West Virginia University, Morgantown, WV 26506, USA . \\ $^{4}$ Also
  adjunct at the National Radio Astronomy Observatory, Green Bank, WV,
  USA. }  \date{1 January 2011} \title[RRATs: Discoveries, Timing \&
  Musings]{RRATs: New Discoveries, Timing Solutions \& Musings}

\begin{document}

\maketitle

\begin{abstract}
  We describe observations of Rotating RAdio Transients (RRATs) that
  were discovered in a re-analysis of the Parkes Multi-beam Pulsar
  Survey (PMPS). The sources have now been monitored for sufficiently
  long to obtain seven new coherent timing solutions, to make a total
  of 14 now known. Furthermore we announce the discovery of 7 new
  transient sources, one of which may be extragalactic in origin (with
  $z\sim0.1$) and would then be a second example of the so-called
  `Lorimer burst'. The timing solutions allow us to infer neutron star
  characteristics such as energy-loss rate, magnetic field strength
  and evolutionary timescales, as well as facilitating
  multi-wavelength followup by providing accurate astrometry. All of
  this enables us to consider the question of whether or not RRATs are
  in any way special, i.e. a distinct and separate population of
  neutron stars, as has been previously suggested. We see no reason to
  consider `RRAT' as anything other than a detection label, the
  subject of a selection effect in the parameter space searched.
  However, single-pulse searches can be utilised to great effect to
  identify pulsars difficult, or impossible, to find by other means,
  in particular those with long-periods (half of the PMPS RRATs have
  periods greater than 4 seconds), high-magnetic field strengths
  ($B\gtrsim 10^{13}$~G) and pulsars approaching the `death
  valley'. The detailed nulling properties of such pulsars are unknown
  but the mounting evidence suggests a broad range of behaviour in the
  pulsar population. The group of RRATs fit in to the picture where
  pulsar magnetospheres switch between stable configurations.
\end{abstract}


\section{Introduction}

A recent highlight in radio transient searches has been the discovery
of RRATs (Rotating RAdio Tranients) by \citet{mll+06} (M+06 from
herein). RRATs have primarily been studied at radio frequencies of
1.4~GHz, where they exhibit detectable emission only sporadically,
with millisecond-duration bursts every few minutes to hours. They are
believed to be neutron stars, for a number of reasons: (1) Causality
implies that pulses of width $W$ originate from emission regions with
size $\leq300\;\mathrm{km}(W/1\;\mathrm{ms})$, which in the case of
RRATs (pulse widths of $\sim1-30$~ms) is much smaller than typical
white dwarf radii. RRAT pulse widths are also similar to the single
pulse widths of radio pulsars (see e.g.~\citet{lk05}). Furthermore,
the dynamical time $t_{\mathrm{dyn}}=1/\sqrt{G\rho}$, where $G$ is
Newton's constant and $\rho$ is mass density, is the scale on which we
expect to see changes, so that the millisecond radio sky consists
mainly of neutron stars which have $t_{\mathrm{dyn, NS}}\sim 0.1$~ms,
whereas transient emission can be expected from white dwarfs on longer
timescales of $t_{\mathrm{dyn, WD}}\sim1-10$~s; (2) Their pulses have
high brightness temperatures of $\sim10^{20}-10^{24}$~K, similar to
radio pulsars (see Figure~\ref{fig:phase_space}); (3) Their underlying
periodicities span the range $0.1-7.7$~s, typical neutron star
rotation periods; (4) One source, J1819$-$1458 has been observed in
the X-ray, showing thermal emission at $\sim140$~eV, as expected for a
cooling neutron star~\citep{rbg+06,mrg+07,rmg+09};
(5) In those sources which have been well studied, their periods are
seen to increase at rates similar to those seen in other neutron star
classes~\citep{mll+06,mlk+09}.

It was initially thought that the RRATs may constitute a heretofore
unknown, distinct population of Galactic neutron stars. However, this
seems unlikely, as when the large projected population of RRATs is
incorporated into the menagerie of other known neutron star classes, a
problem results. If the known neutron star groups are distinct, then
the Galactic supernova rate is insufficient to explain the number of
neutron stars which we infer~\citep{kk08}. This problem can be removed
if the groups of neutron stars are evolutionarily linked and/or if
their projected populations are over-estimated. An evolutionary link
between the various classes would, in some senses, be satisfactory, as
such links must exist. However neutron star spin evolution is poorly
understood, and no such evolutionary framework exists (see
\citet{vm10} for a recent discussion of this). A large over-estimate
of the population is also possible, given the large extrapolation from
a small number of known objects to an entire Galactic population.

Such motives resulted in our re-analysis of an archival pulsar survey,
resulting in the discovery of 11 new RRATs, which we described in
\citet{kle+10} (K+10 from herein), and several other authors have
performed successful searches also (see \S~\ref{sec:discussion} which
provides a census of known sources). We describe, in
\S~\ref{sec:observations_analysis}, the methods used, and difficulties
encountered, in obtaining coherent timing solutions for these
sources. As well as identifying some further discoveries in
\S~\ref{sec:new_discoveries}, we present followup observations of the
new RRATs we previously identified. We have been able to obtain
solutions for seven sources, described in \S~\ref{sec:new_solutions},
which doubles the number of RRAT timing solutions that are known. We
discuss the importance of timing solutions, including what they allow
us to infer about the neutron stars, the ability to monitor for glitch
activity (which has been seen but whose significance is yet to be
appreciated; see \S~\ref{sec:switching_magnetospheres}), and important
benefits such as vastly improved astrometry. In
\S~\ref{sec:discussion} we review what is now known about neutron
stars detected as RRATs, and consider the question of whether they are
in any way distinct from radio pulsars, before making our conclusions
in \S~\ref{sec:conclusion}.

\section{Observations \& Analysis}\label{sec:observations_analysis}

\subsection{The PMSingle Analysis}
\begin{figure*}
  \begin{center}
    \includegraphics[trim = 23mm 20mm 0mm 10mm, clip, scale=0.56]{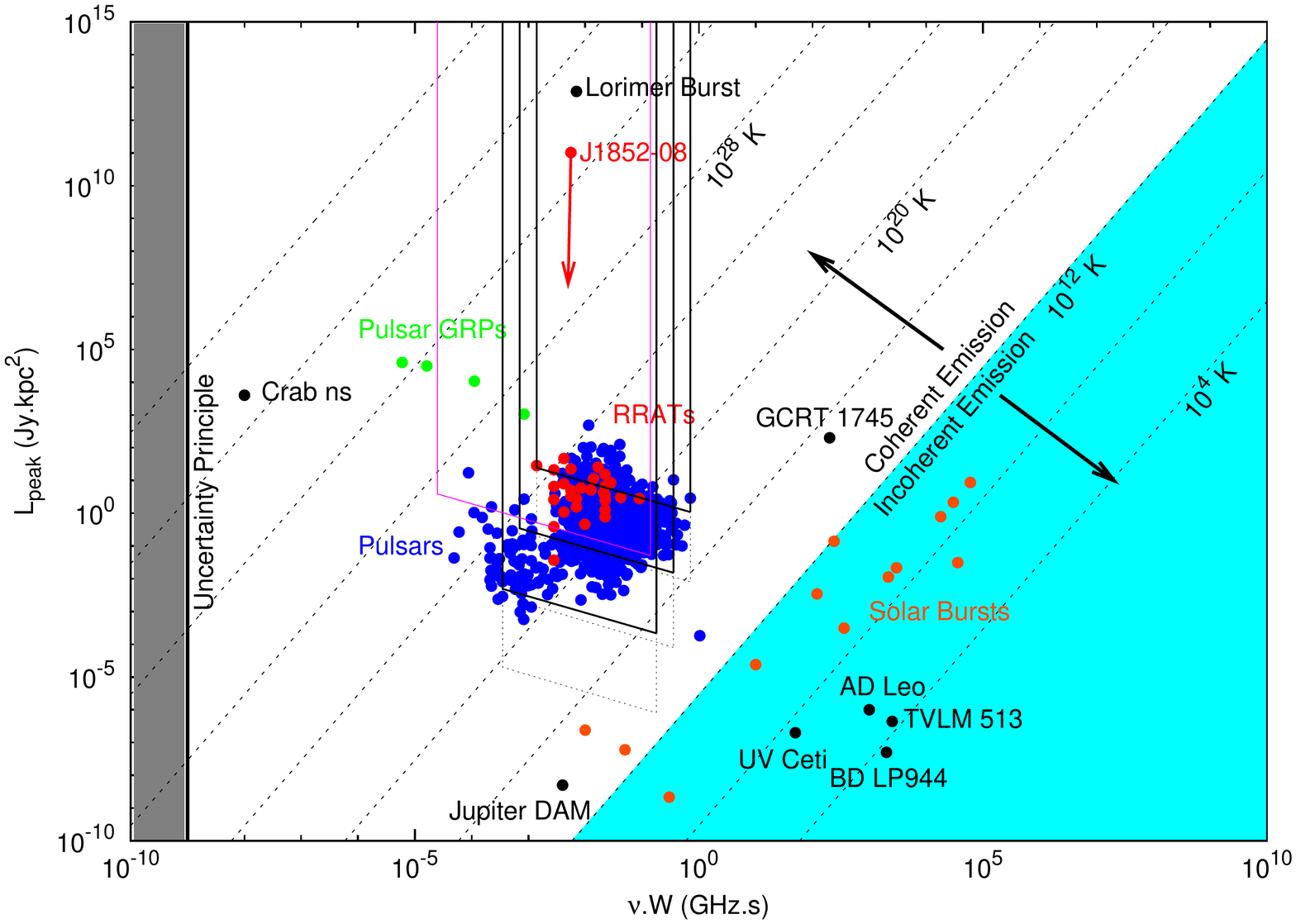}
  \end{center}
  \vspace{-5pt}
  \caption{\small{The transient `phase space' with known sources
      identified. This is simply a plot of the radio
      (pseudo-)luminosity $L=SD^2$ versus $\nu W$, where $S$ is flux
      density, $D$ is distance, $\nu$ is observing frequency and $W$
      is pulse width. As radio frequencies are in the Rayleigh-Jeans
      regime ($h\nu \ll kT$) we can draw lines of constant
      \textit{minimum} brightness temperature
      $T_{\mathrm{B}}=4\times10^{17}(SD^2/\mathrm{Jy\,kpc}^2)
      (\mathrm{GHz\,s/\nu W})^2$ (see \citet{kea10a} or
      \citet{kea10}). Plotted are pulsars~\citep{hmth04a}, `RRATs',
      pulsar `giant radio pulses'~\citep{cstt96,rj01}, flare
      stars~\citep{bastian94,rwgr03,ob08,osten08}, auroral radio
      emission from the Sun and planets~\citep{dulk85,zarka98},
      GCRT~1745$-$3009~\citep{hlr+06} and the so-called `Lorimer
      burst'~\citep{lbm+07}, which we give only as a representative
      but not exhaustive list of sources. The boundary between
      coherent and incoherent emission is at $\approx10^{12}$~K, due
      to inverse Compton cooling~\citep{red94}. The sensitivity of the
      PMSingle analysis (black lines) to individual bursts, is
      overplotted, from lowest to highest $L$, for distances of $0.1$,
      $1$ and $10$~kpc respectively. With the effective area of the
      SKA the curves become lower by $\gtrsim2$ orders of magnitude in
      $L$ (dotted lines). The LOFAR survey sensitivity curve (pink
      line) for a distance of $2$~kpc is also shown.}}
  \vspace{-5pt}
  \label{fig:phase_space}
\end{figure*}

In K+10 we described our reprocessing (which we refer to as PMSingle)
of the Parkes Multi-beam Pulsar Survey (PMPS), a survey of the
Galactic plane between $l=260\degree$ and $l=50\degree$, and
$|b|<5\degree$. The survey was performed using $96\times3$~MHz
frequency channels centred at an observing frequency of 1374~MHz, with
250-$\mu$s time sampling. Detailed survey specifics can be found in
\citet{mlc+01}.  The analysis described in K+10 resulted in the
discovery of 11 new RRAT sources. We have now made an additional
confirmation (see \S~\ref{sec:new_discoveries_pmps}), so that there
are now 12 sources, discovered in the PMSingle analysis, which have
been detected on multiple occasions. These sources have been the
subject of an ongoing campaign of observations which we describe below
in \S~\ref{sec:bits_to_bats} and \S~\ref{sec:new_solutions}. In
addition to these repeating sources, we have identfied seven sources
which have not been re-detected since their discovery
observations. Nonetheless we consider the astrophysical nature of
these sources to be self-evident, as we describe below in
\S~\ref{sec:isolated} and \S~\ref{sec:1852}. These 19 sources, added
to the 11 identified in \citet{mll+06} mean that there have now been
30 such transient radio sources discovered in the PMPS. The detection
statistics of the PMSingle discoveries are given in
Table~\ref{tab:pmsingle_rrat_properties}, and
Figure~\ref{fig:phase_space} shows where these sources lie in the
``transient phase space'' defined by \citet{clm04}.

\begin{table*}
  \begin{center}
    \caption{\small{The observed properties of the newly identified
        sources from the PMSingle analysis, as well as those BB10
        candidates which we have confirmed. Note that for sources
        previously published in K+10 the DM values are no more precise
        due to a lack of multi-frequency observations, necessary for
        accurately determining DM. The distances quoted are those
        derived from the DM using the NE2001 model of the electron
        content of the Galaxy~\citep{cl02}, with typical errors of 20
        percent. The quoted 1.4-GHz peak flux densities are determined
        by using the radiometer equation~\citep{lk05} and using the
        known gain and system temperature of the 20-cm multi-beam
        receiver (as given in the April 6, 2009 version of the Parkes
        Radio Telescope Users Guide) and have typical uncertainties of
        30 percent level. The $\dagger$ denotes the fact that
        J1652$-$44 has been detected in just 1 of 28 observations as a
        single pulse source. It is detected in 22 of these
        observations as a folded source. The $\ddagger$ denotes two
        sources for which a detection has recently been reported in
        the HTRU
        survey~\citep{bbj+11}}}\label{tab:pmsingle_rrat_properties}
    \begin{tabular}{ccccccccccc}
      \hline\hline Source & DM & D & P & w & $S_{\rm{peak}}$ & $L_{\rm{peak}}$ & $N_{\mathrm{det}}/N_{\mathrm{obs}}$ & $N_{\mathrm{pulses}}$ & $T_{\mathrm{obs}}$ & $\dot{\chi}$ \\
      & ($\rm{cm^{-3}\,pc}$) & (kpc) & (s) & (ms) & (mJy) & ($\rm{Jy\,kpc^2}$) & & & (hr) & (hr$^{-1}$) \\
      \hline
      \multicolumn{3}{l}{\textbf{Repeating sources}} \\
      J1047$-$58 & 69.3(3.3) & 2.3 & 1.231 & 4 & 630 & 3.3 & 7/28 & 60 & 16.0 & 3.8 \\
      J1423$-$56$\ddagger$ & 32.9(1.1) & 1.3 & 1.427 & 5 & 930 & 1.5 & 13/22 & 48 & 15.0 & 3.2 \\
      J1514$-$59 & 171.7(0.9) & 3.1 & 1.046 & 3 & 830 & 7.9 & 32/32 & 361 & 19.2 & 18.7 \\
      J1554$-$52 & 130.8(0.3) & 4.5 &  0.125 & 1 & 1400 & 28.3 & 37/37 & 703 & 13.5 & 52.0 \\
      J1652$-$44 & 786(10.0) & 8.4 & 7.707 & 64 & 40 & 2.9 & 1/28$\dagger$ & 9 & 13.1 & 0.7 \\
      J1703$-$38 & 375(12.0) & 5.7 & - & 9 & 160 & 5.1 & 13/18 & 25 & 14.1 & 1.7 \\
      J1707$-$44 & 380(10.0) & 6.7 & 5.764 & 12 & 575 & 25.8 & 26/27 & 129 & 14.4 & 8.9 \\
      J1724$-$35 & 554.9(9.9) & 5.7 & 1.422 & 6 & 180 & 5.8 & 17/23 & 49 & 14.9 & 3.2 \\
      J1727$-$29 & 92.8(9.4) & 1.7 & - & 7 & 160 & 0.4 & 4/11 & 4 & 6.1 & 0.6 \\
      J1807$-$25 & 385(10.0) & 7.4 & 2.764 & 4 & 410 & 22.4 & 25/25 & 149 & 18.1 & 8.2 \\
      J1841$-$14 & 19.4(1.4) & 0.8 & 6.598 & 3 & 1700 & 1.0 & 42/43 & 989 & 15.6 & 63.4 \\
      J1854$+$03 & 192.4(5.2) & 5.3 & 4.558 & 16 & 540 & 15.1 & 29/32 & 146 & 16.3 & 8.9 \\

      \multicolumn{3}{l}{\textbf{Non-repeating sources}} \\
      J0845$-$36 & 29(2) & 0.4 & - & 2 & 230 & 0.04 & 1/2 & 1 & 1.1 & 1.8 \\
      J1111$-$55 & 235(5) & 5.6 & - & 16 & 80 & 2.5 & 1/7 & 2 & 4.3 & 0.4 \\
      J1308$-$67$\ddagger$ & 44(2) & 1.2 & - & 2 & 270 & 0.4 & 1/5 & 2 & 3.1 & 0.6 \\
      J1311$-$59 & 152(5) & 3.1 & - & 16 & 130 & 1.3 & 1/6 & 1 & 3.3 & 0.3 \\
      J1404$-$58 & 229(5) & 4.8 & - & 4 & 220 & 5.1 & 1/10 & 7 & 6.1 & 1.1 \\
      J1649$-$46 & 394(10) & 5.1 & - & 16 & 135 & 3.5 & 1/4 & 1 & 3.4 & 0.3 \\
      J1852$-$08 & 745(10) & $\sim500000$ & - & 7 & 410 & $\sim10^7$ & 1/9 & 1 & 4.2 & 0.2 \\

      \multicolumn{3}{l}{\textbf{BB10 RRATs}} \\                                                                                   
      J0735$-$62 & 19(8) & 0.9 & 4.865 & 2 & 580 & 0.5 & 3/6 & 36 & 1.9 & 18.9  \\
      J1226$-$32 & 37(10) & 1.4 & 6.193 & 12 & 590 & 1.2 & 13/20 & 112 & 5.8 & 19.2  \\
      J1654$-$23 & 74.5(2.5) & 2.0 & 0.545 & 1 & 1300 & 5.2 & 13/14 & 151 & 2.6 & 57.9 \\

      \hline
    \end{tabular}
  \end{center}
\end{table*}

\subsection{Pulsar Timing}\label{sec:psr_timing}

Pulsars are commonly referred to as stable astrophysical
clocks. However, even though they are rotationally stable, on a
period-by-period basis the pulses we detect from pulsars are variable
in amplitude, phase and shape. These individual pulses (aka
sub-pulses) can vary in random, as well as highly ordered,
ways. Sub-pulse drifting is a phenomenon whereby the rotational phase
wherein we see pulsar emission changes periodically (see
e.g.~\citet{wes06}). Some pulsars also exhibit `mode-changing', or
`moding', whereby they are seen to switch between two or more
different stable emission profiles~\citep{bmsh82}. Another phenomenon
is nulling, which can be seen as an extreme example of moding, where
one of the modes shows no radio emission, i.e. the radio emission
ceases and the pulsar is `off'~(e.g. \citet{bac70}). Random changes
are usually labelled as `pulse jitter', e.g. the Gaussian variations
in pulse phase seen in PSR~J0437$-$4715~\citep{cs10}. We will discuss
these phenomena again in \S~\ref{sec:discussion}. For the purposes of
`timing' a pulsar, i.e. modelling its rotational phase as a function
of time with respect to pulsar and astrometric parameters, these
variations all amount to `timing instabilities'. We note that none of
these effects are symptomatic of \textit{rotational} irregularities
--- the pulsar is still spinning down in a well-behaved manner. What
is variable/unstable is the source of the radio emission. There are
also rotational instabilities known as glitches which are single
events consisting of instantaneous jumps in rotation frequency and its
derivatives (see e.g.~\citet{esp09,elsk11}). Additionally, the
possibly more general phenomena of slow-down rate switching may be
occurring in much of the pulsar population~\citep{lhk+10}, something
which we consider further in \S~\ref{sec:discussion}.

\subsection{Integrated Profiles}\label{sec:integrated_profiles}
To perform `pulsar timing' of a source it is usually observed for a
large number of contiguous pulse periods, which are integrated to
create an average pulse profile $P(t)$. The addition of many pulse
periods is performed for two reasons: (1) to compensate for all of the
timing instabilities outlined above, and (2) to increase the
signal-to-noise ratio of $P(t)$. We note that a high signal-to-noise
ratio does not imply a stable profile (we define stability below). In
practise, as many periods as possible are used in timing
`normal'/`slow' pulsars, typically $10^2-10^3$, but for the faster
millisecond pulsars (MSPs) $\gtrsim 10^5$ are used
routinely. Determining a pulse time-of-arrival (TOA) for a given
observation then amounts to cross-correlating the observed profile
$P(t)$ with a very high S/N (or sometimes even analytic) template
profile $T(t)$ under the \textit{assumption} that the profile is just
a shifted, scaled and noisier version of the template, i.e.
\begin{equation}\label{eq:prof}
  P(t)=AT(t+\psi)+N(t)\;,
\end{equation}
where $A$ is a scale factor, $\psi$ is a phase shift and $N$ is an
additive noise term~\citep{tay90a}. Determining $\psi$ gives the TOA
relative to some known reference time, usually the observatory clock.

Equation~\eqref{eq:prof} is valid if the profile is stable. For a
profile to be stable its correlation coefficient with the template,
$R=R(n)$, must improve according to $\langle 1-R(n) \rangle \propto
n^{-1}$ where $n$ is the number of periods averaged over to make the
template (Liu et al., in preparation). In practise this is realised
only after we have averaged some critical number of periods to make a
profile. For smaller values of $n$, $\langle 1-R \rangle$ will improve
faster than $n^{-1}$. Breaks in $\langle 1-R \rangle $ at certain
values of $n$ indicate periodic instabilities, e.g. drifting and
nulling timescales. Beyond some value $n_{\mathrm{crit}}$, when
$\langle 1-R(n)\rangle \propto n^{-1}$ we say that $P(t)$ is
stable. We note that it has, in the past, been suggested that $\langle
1-R(n) \rangle \propto n^{-0.5}$ signalled
stability~\citep{hmt75,rr95,lk05}\footnote{Furthermore, in the past,
  arbitrary criteria for `stability' have been set, e.g.~\citet{hmt75}
  defined stability as $R=0.9995$.} but this is incorrect. For MSPs,
this critical number of periods is $\lesssim10^4$ and is always
reached so that precision timing can be performed. In the case of
slower pulsars the stability criterion is not
reached~\citep{hmt75,rr95}, nor is the precision as high given that
the TOA error $\sigma_{\rm{TOA}}\propto W^{3/2}P^{-1/2}$ is larger for
slow pulsars than for MSPs, where $W$ and $P$ are the pulse width and
period, respectively. Furthermore the slower pulsars are observed to
exhibit more glitches and more so-called `timing
noise'\footnote{Timing noise is a red noise feature seen in pulsar
  timing residuals which may be related to pulsars switching between
  two spin-down rates~\citep{lhk+10}.}. Thus MSPs can be timed with
very high precision whereas slow pulsars cannot.

\subsection{Single Pulses}
RRATs are generally detected via their sporadic single pulses as (by
definition) they are only, or more easily, detectable in this way as
opposed to methods relying on time-averaged flux. Their pulses are not
detectable every rotation period and the typical observed
pulse-to-pulse separations range from $\sim10$ to $\sim1000$ periods
so that, unlike typical pulsars, we do not see strong pulse profiles
after folding. This means we lose the two advantages of phase folding
--- stable profiles and increased signal-to-noise ratio. However the
single pulses themselves are quite strong with typical peak flux
densities of $\sim10^2-10^3$~mJy (see
Table~\ref{tab:pmsingle_rrat_properties}) and for the observations
reported here the typical signal-to-noise ratios this corresponds to a
range from as low as 6 to as high as 60 so that, from a signal
intensity point of view, timing RRATs from their single pulses is
possible. However, the single pulse profiles are far from stable in
phase. Phase stability is usually implicitly assumed (in timing
analysis software) when using high S/N profiles and templates. This
assumption is inappropriate for single-pulse timing (as it is for slow
pulsars timed using unstable average profiles) and will result in
extra scatter in our timing residuals with a magnitude given by the
size of the phase window wherein we see single pulses. As we will show
this effect is clear in our data (see
Figure~\ref{fig:pmsingle_resids}, as well as Figure 1a of
\citet{lmk+09}).

\subsection{Observations \& Timing}\label{sec:bits_to_bats}


Here we outline the steps involved in progressing from a telescope
signal to barycentred pulse arrival times and a coherent timing
solution.

(i) \textit{Observe sources in `search mode'.} The followup
observations at Parkes consist of sporadic observations between
October 2008 and March 2009, and regular approximately monthly
observations since April 2009, which are ongoing. In our observational
setup we utilise a bandwidth of 256~MHz divided into 512 channels,
sampled every 100~$\mu$s. The telescope receives dual linear
polarisations but these are summed to produce total intensity,
i.e. Stokes I. The data are 1-bit digitised before being written to
tape. 
The beginning of the observation is time-stamped according to the
observatory clock, 
and is known to an accuracy of $\sim80$~ns.

(ii) \textit{Search the data for single pulses}. As described in K+10
the data are searched for strong, dispersed single pulses of
radiation. 
Once detected, dedispersed single pulse profiles, are extracted from
the data.



(iii) \textit{Obtain TOAs.} The templates used here are empirical and
derived from smoothing each source's strongest observed pulse which
results in simple one component templates. Averaging all of the
(detected) individual pulses gives a wider pulse profile unsuitable
for cross-correlating with individual pulses. The profiles are
cross-correlated with the template and $\psi$ determined to obtain the
TOA at the telescope, i.e. the site arrival time (SAT, aka topocentric
arrival time), which is referenced to the time stamp.

(iv) \textit{Convert SATs to BATs}. SATs are measured in Coordinated
Universal Time (UTC). These are converted to barycentric arrival times
(BATs), i.e. arrival times at the solar system barycentre at infinite
frequency (with dispersive delay removed) in Barycentric Coordinate
Time (TCB). The steps involved in this conversion, and definitions of
these time systems are well described elsewhere (Lorimer \& Kramer
2005, Hobbs et al. 2006; Edwards et al. 2006, or see Appendix E of
Keane 2010)\nocite{lk05,hem06,ehm06,kea10a}.

Once we have obtained BATs we can model the timing parameters of the
source. This is done using
\textsc{psrtime}\footnote{http://www.jb.man.ac.uk/$\sim$pulsar/observing/progs/psrtime.html}
and
\textsc{tempo2}\footnote{http://www.atnf.csiro.au/research/pulsar/tempo2/},
standard pulsar timing software packages. If we express the rotational
frequency of the pulsar as a Taylor expansion
\begin{equation}\label{eq:nu_taylor}
  \nu(t)=\nu_0+\dot{\nu}_0(t-t_0)+\frac{1}{2}\ddot{\nu}_0(t-t_0)^2+\ldots
  \; ,
\end{equation}
the rotational phase (simply the integral of frequency with respect to
time, modulo $2\pi$) is given by
\begin{align}\label{eq:timing_phase}
  \phi(t) =  \phi_0 &+ \nu_0(t-t_0)+\frac{1}{2}\dot{\nu}_0(t-t_0)^2
  \nonumber
  \\ & +\frac{1}{6}\ddot{\nu}_0(t-t_0)^3+\ldots\;\;\;\;\;\;\;\;\;\;\;(\rm{mod\;2\pi}).
\end{align}
In addition to these terms, binary effects should be added (however
none of the sources discussed here have detected binary companions)
and the \textit{observed} phase will be different due to positional
uncertainties. 
Timing consists of minimising the $\chi^2$ of the residuals of our
timing model, i.e. the difference between our model for when pulses
arrive and when they actually arrive (the BATs we measure).

Immediately after discovering and confirming a new source we know very
little about it. If the rate of pulse detection is too low then we
will not be able to determine an estimate of the period using period
differencing. In this case there is no way to proceed with timing the
source. Assuming the rate is sufficient then we have an initial guess
for the period and a knowledge of the sky position (uncertain to
$\sim7$~arcmin in both right ascension and declination, corresponding
to the beamwidth of a pointing in the PMPS) which serves as our
initial guess of the timing ephemeris. We can see from
Equation~\ref{eq:timing_phase} that different effects will become
visible in our residuals over different timescales. On the shortest
timescale all we need to worry about is the rotation frequency,
$\nu$. We begin our timing solution by obtaining several closely
spaced `timing points', (say) every 8 hours over the space of a day or
two. This is necessary to build a coherent solution on short
timescales as our initial knowledge of the period is not sufficient to
be able to combine, in phase, TOAs obtained a few days apart.
Once this has been done the period will be known to sufficient
accuracy that all our TOAs over the timescale of a few days will be in
phase. If we monitor the source like this 
we will notice a quadratic signature appear in our residuals. This is
the effect of the frequency derivative $\dot{\nu}$ (which is initially
set to zero). For the sources reported here, this $\dot{\nu}$ effect
is seen over a timescale of weeks to months. Positional uncertainties
result in sinusoids, with periods of one year, appearing in the
residuals. If the sky position is not well known, it is difficult to
disentangle the effects of spin-down rate and positional uncertainty
until at least 6~months of monitoring has been made, and preferably at
least one year (i.e. a quadratic curve is highly covariant with half a
sine wave).


\section{New Discoveries}\label{sec:new_discoveries}

\subsection{J1652$-$44}\label{sec:new_discoveries_pmps}

\begin{figure}
  \begin{center}
    \includegraphics[trim = 20mm 0mm 0mm 0mm, clip, scale=0.33,angle=0]{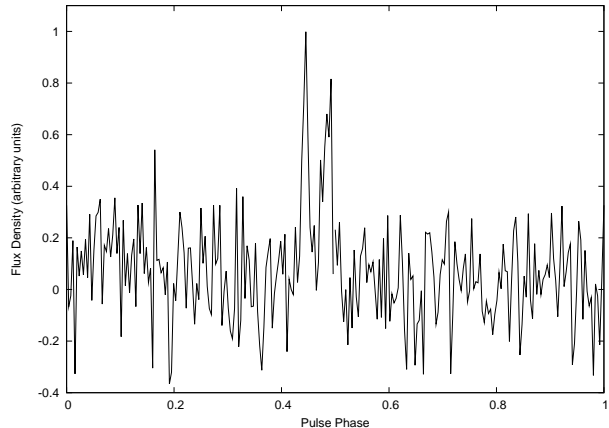}
  \end{center}
  \vspace{-25pt}
  \caption{\small{A typical pulse profile from a 30-minute observation
      of J1652$-$44.}}
  \label{fig:1652_pdmp}
\end{figure}

J1652$-$44 was one of the Class 1 candidates found in the analyses
presented in K+10. Despite showing 9 strong pulses in its discovery
observation, confirmation was difficult. A small number of bursts have
since been observed but none as strong as in the original survey
observation. These borderline detections were not enough to
conclusively confirm the candidate but it turned out that J1652$-$44
was sometimes detectable by folding the time series at the period of
the pulsar. Looking for a folded signal was made possible by obtaining
an initial period of $P=7.70718$~s from period differencing of the
discovery burst times of arrival (TOAs, see K+10 for a description of
this). Using this, and the dispersion measure (DM) at which the bursts
peaked, as a starting point, each of 28 followup observations were
folded and dedispersed into archives consisting of 1-minute
subintegrations. A search in period and DM was then performed using
\textsc{pdmp}\footnote{\texttt{http://psrchive.sourceforge.net/manuals/pdmp}}.
In 22 of the observations a folded signal, like that shown in
Figure~\ref{fig:1652_pdmp}, was detected with a double-peaked profile.

\subsection{Single Detections}\label{sec:isolated}
Additionally, we have identified 6 of the PMSingle candidates which we
consider to be `self-confirmed', i.e. we have not re-observed bursts
in our followup observations but we deem the survey detection
sufficiently convincing that the astrophysical nature of these sources
is clear. A number of these are just single bursts, showing the
characteristic dispersive delay expected from celestial sources, are
detected in only one of the 13 beams and show no signatures of
RFI. Figure~\ref{fig:selfconfirmed} shows an indicative frequency
versus time plot demonstrating the dispersive sweep of an individual
pulse. We note that unlike the bursts reported by \citet{bbe+11},
whose origin appears to be terrestrial, no ``kinky'' deviations are
seen from the ideal dispersion law, nor are any of these events
detected in multiple beams.

The sources show between one and seven pulses in their discovery
observations and have been followed up for between one and six hours,
without showing further pulses. As these bursts are just a few
milliseconds in duration a neutron star
is expected to be the source of the emission. As the discovery
observations clearly show these sources to be astrophysical, the long
followups with no confirmation suggest a very low rate of
bursting. They therefore have significant implications for the
population size of such sources. For instance, if a source shows one
burst in 5 hours of observation, it suggests that, as a zeroth order
estimate, $\sim9$ such sources may have been missed during the survey
which consisted of 35-minute pointings. In this sense then, the longer
these sources remain unconfirmed the more interesting they are. For
those sources which have shown just one burst (J0845$-$36, J1311$-$59,
J1649$-$46 and J1852$-$08) we cannot rule out some theoretically
predicted explanations which would not be expected to repeat,
e.g. annihilating mini black holes~\citep{rees77},
supernovae~\citep{pt79} or merging neutron star
binaries~\citep{hl01b}. Observing repeated bursts rules out such
events and points at a temporarily re-activated `dead' pulsar as a
likely origin.

\subsection{J1852$-$08}\label{sec:1852}
\begin{figure}  
  \begin{center}
    \hspace{-2.1mm}\includegraphics[trim = 20mm 54mm 45mm 47mm, clip, scale=0.37275,angle=0]{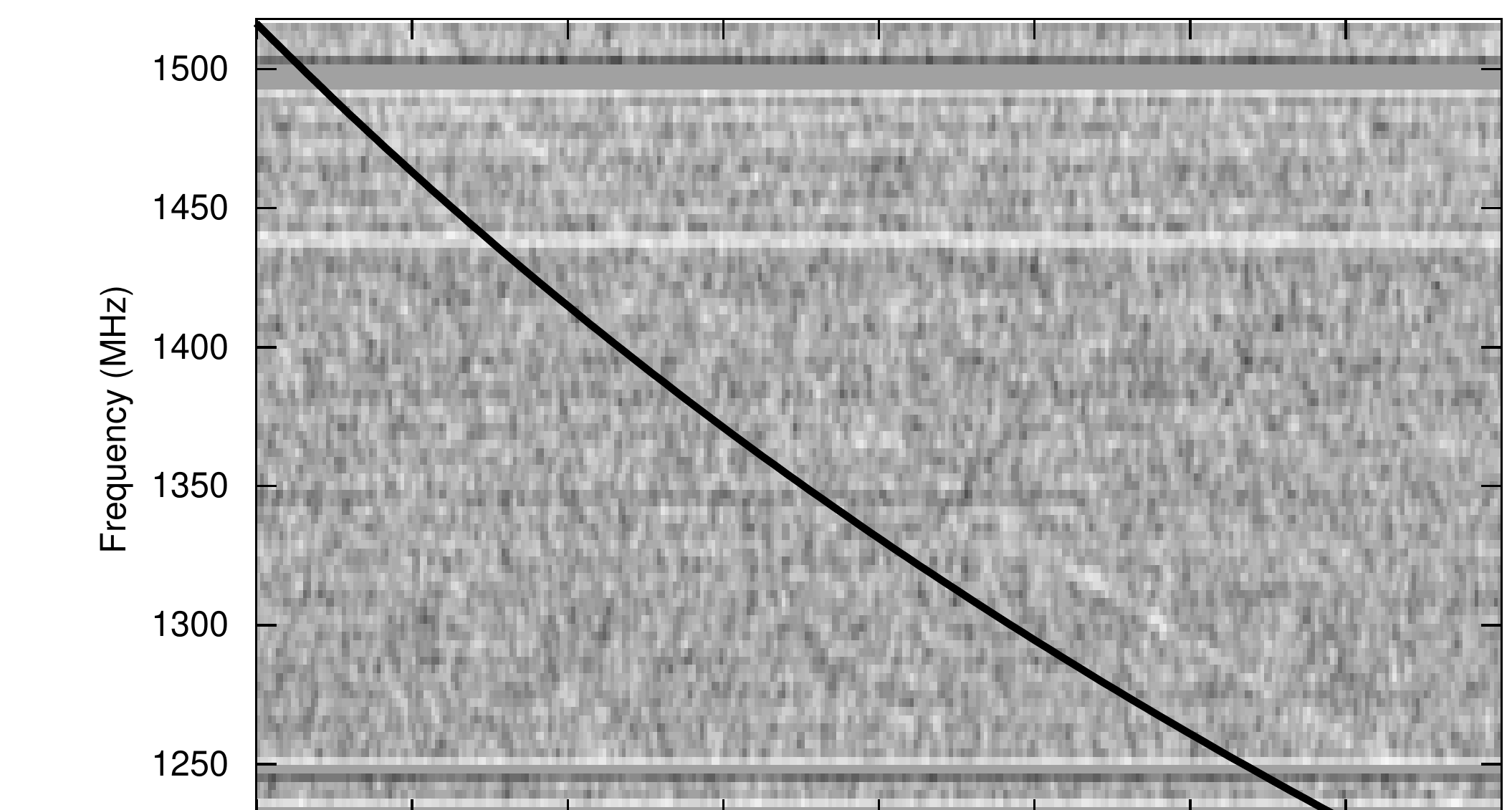} 
    \includegraphics[trim = 0mm 0mm 0mm 27mm, clip, scale=0.305,angle=0]{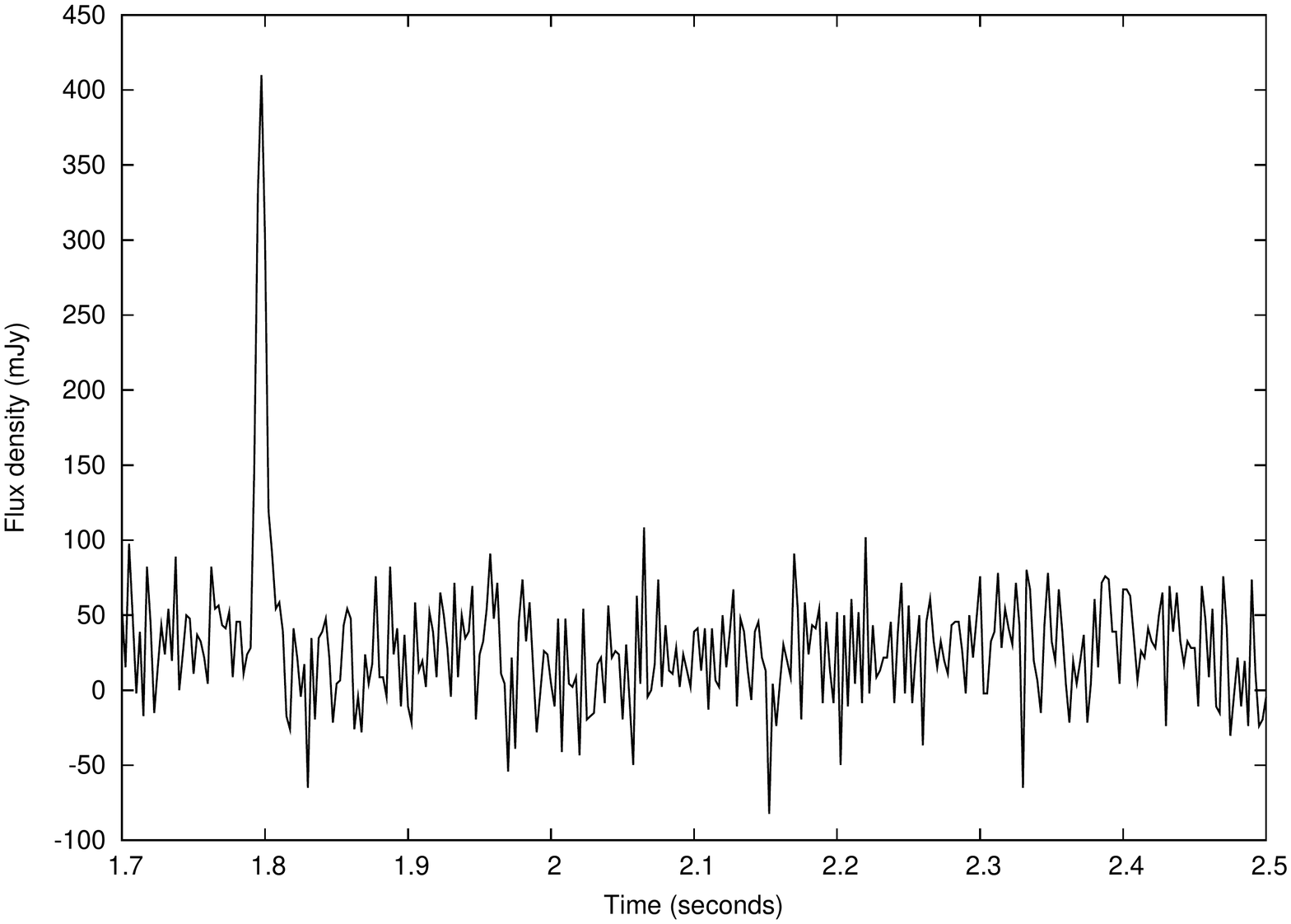}
  \end{center}  
  \vspace{-20pt}
  \caption{\small{(Top) A plot of the J1852$-$08 burst in
      frequency-time space, aka a `dynamic spectrum'. The theoretical
      dispersion law is $ t_{\mathrm{delay}}=4150\;\left(DM/f^2\right)
      \;\mathrm{sec}$, where $DM$ is the dispersion measure in units
      of $\mathrm{cm}^{-3}\,\mathrm{pc}$ and $f$ is the observing
      frequency in MHz. The offset black solid line is the theoretical
      curve for a source with $DM=745\;\mathrm{cm^{-3}\,pc}$, which is
      clearly obeyed by the pulse. (Bottom) The dedispersed pulse,
      i.e. the dynamic spectrum collapsed along a slope given by the
      theoretical curve. The flux density scale is uncertain by up to
      30 percent.}}
  \label{fig:selfconfirmed}
\end{figure}

The most interesting single pulse source is J1852$-$08, an isolated
7-ms pulse with a dispersion measure of $745\;\mathrm{cm^{-3}\,pc}$
(see Figure~\ref{fig:selfconfirmed}). Dividing the band into 8
sub-bands and obtaining TOAs for each shows a frequency
dependent-delay between each TOA of the form $f^{-\alpha}$ where
$\alpha=2.02(1)$, consistent with the theoretical value of $2$ for a
cold ionised inter-stellar medium. Hence the pulse is unlikely to be
due to a terrestrial source.
We note that the half-amplitude pulse width is slightly wider in the
bottom half of the band, at $9.1$~ms compared to $7.1$~ms in the top
half of the band, although there is no obvious indication, permitted
by the signal-to-noise ratio, of scattering, e.g. an exponential tail,
so that this may be intrinsic to the pulse. Dedispersing the entire
band gives a pulse width of $7.3$~ms. We note that the empirical model
of \citet{bcc+04} predicts a scattering time of $\sim 130$~ms, which
is not seen here, where the scattering time can be no more than a few
milliseconds. However this empirically determined relation between
scattering, DM and observing frequency has observed deviations of more
than an order of magnitude in either direction.

The Galactic coordinates of this source are $l=25.4\degree$,
$b=-4.0\degree$, so that this large DM implies an extragalactic
distance for this source. According to the NE2001 Galactic electron
density model~\citep{cl02} the maximum contribution from the Galaxy
along this line of sight is
$DM_{\mathrm{Gal}}=533\;\mathrm{cm}^{-3}\,\mathrm{pc}$. Thus a
Galactic explanation of this source requires that the NE2001 model is
incorrect along this line of sight. If there were an unknown
contribution to the free electron density, $DM_{\mathrm{Gal}}$ would
increase, and the inferred distance to J1852$-$08 could be drastically
reduced (see \citet{dtbr09} for a discussion of errors in NE2001
distances). In that case the burst from J1852$-$08 would appear to be
a giant radio pulse from a pulsar. However for typical giant radio
pulse distributions (see e.g. \citet{kss10}) this means that we would
already have detected many weaker pulses, which is not the case.

If NE2001 is reliable along this line of sight then the surplus of
$DM_{\mathrm{extra}}=222\;\mathrm{cm}^{-3}\,\mathrm{pc}$ must be due
to extragalactic contributions (the inter-galactic medium and any
putative host galaxy). A DM-redshift relation is known~\citep{ioka03}
which would apply to this component, and takes the form:
$DM_{\mathrm{extra}}\approx 1200z\;\mathrm{cm}^{-3}\,\mathrm{pc}$. If
all of the $DM_{\mathrm{extra}}$ component is due to the
inter-galactic medium, the inferred redshift and distance are
$z\approx0.18$ and $D\approx520h^{-1}$~Mpc ($\Omega_{\mathrm{m}}=0.3$,
$\Omega_{\Lambda}=0.7$). Allowing for a contribution of
$100\;\mathrm{cm}^{-3}\,\mathrm{pc}$ from a host galaxy (see
\citet{lbm+07}), these values become $z\approx0.09$ and
$D\approx260h^{-1}$~Mpc.
This implies the strongest intrinsic peak luminosity of all the PMPS
sources, of $\gtrsim 10^{11}\;\mathrm{Jy\,kpc^2}$. It is noticeable as
the RRAT with the highest peak luminosity in
Figure~\ref{fig:phase_space} where it lies just below the burst
reported by \citet{lbm+07} in transient phase space, and some 6 orders
of magnitude above the giant pulses seen in some radio pulsars. The
SIMBAD\footnote{\texttt{http://simbad.u-strasbg.fr/simbad/}} database
lists no apparent host galaxies for this object, although the
positional uncertainty for this event is quite large, at $\sim7$
arcmin. Furthermore, as this event occured in June 2001, just like the
Lorimer burst, which occured two months later, this burst was a
pre-LIGO and pre-GEO600 event, so no gravitational wave emitting
counterpart can be searched for. We can say that, by causality and the
pulse duration, the source is limited to a maximum size of
2100~km. Thus, if the NE2001 model is correct along this line of
sight, J1852$-$08 fits many of the criteria for being a second example
of the Lorimer burst.

\subsection{Repeating Sources (BB10)}
In addition to the PMPS, two further pulsar surveys have been
performed at the same Galactic longitudes, but at intermediate and
high Galactic latitudes of $5\degree < |b| <
30\degree$~\citep{ebsb01,jbo+09}. These surveys used the same
specifications as the PMPS, except with a faster time sampling of
$125\;\mu$s and shorter pointings of $4.4$~minutes.
Recently,~\citet{bb10}(BB10 from herein) have analysed these surveys
and presented 14 new transient sources, 7 of which were candidates
which had never been confirmed. One of these unconfirmed sources was
in fact re-detected by the authors soon after their publication
(Burke-Spolaor, private communication), but 6 remained unconfirmed. As
part of our observing programme, these 6 sources were observed in
search of single pulses and 3 of these have now been confirmed. Two of
these sources have been regularly observed since January 2010 and both
have provisional timing solutions, which we describe further in
\S~\ref{sec:preliminary_timing}.

\section{New Timing Solutions}\label{sec:new_solutions}
After discovery of a new source, very little is known: a dispersion
measure, a crude knowledge of the position, and perhaps a period.
Determining a timing solution increases our knowledge greatly. It
enables us to infer properties of the star, such as energy-loss rate,
magnetic field strength and evolutionary timescales (see
e.g. \citet{ls04}). We report these values for the PMPS RRATs in
Table~\ref{tab:inferred_properties} and
Figure~\ref{fig:all_pmps_rrat_properties}. Timing solutions also tell
us where in $P-\dot{P}$ space our sources occupy, allowing us to
investigate and/or infer pulsar evolutionary paths. Furthermore we can
identify additional contributions to pulsar spin evolution, in
particular due to glitches, which manifest as step changes in spin
frequency and its derivatives (see e.g. \citet{elsk11}). The accurate
astrometry provided by the timing solutions allow multi-wavelength
observations (see e.g. \citet{dkm+11}), impossible with the poor
spatial resolution of single-dish radio telescopes.
Another spin-off is that an improved retrospective search for pulses
will be possible, allowing an optimal nulling analysis.

Here we report the complete timing solutions for seven PMSingle
sources. These solutions consist of fits in period, period derivative,
right ascension and declination. Figure~\ref{fig:pmsingle_resids}
shows the timing residuals for 6 of these sources (those timed via
their single pulses, all but J1652$-$4406) and
Table~\ref{tab:pmsingle_timing_solutions} gives the parameters of the
fits. Below we quickly review each of the sources in turn, before
giving updates on provisional timing solutions of PMSingle and BB10
sources which do not yet have a timing solution.

\subsection{Complete Timing Solutions}

\begin{figure*}
  \begin{center}
    \includegraphics[trim = 20mm 15mm 165mm 18mm, clip, scale=0.4,angle=-90]{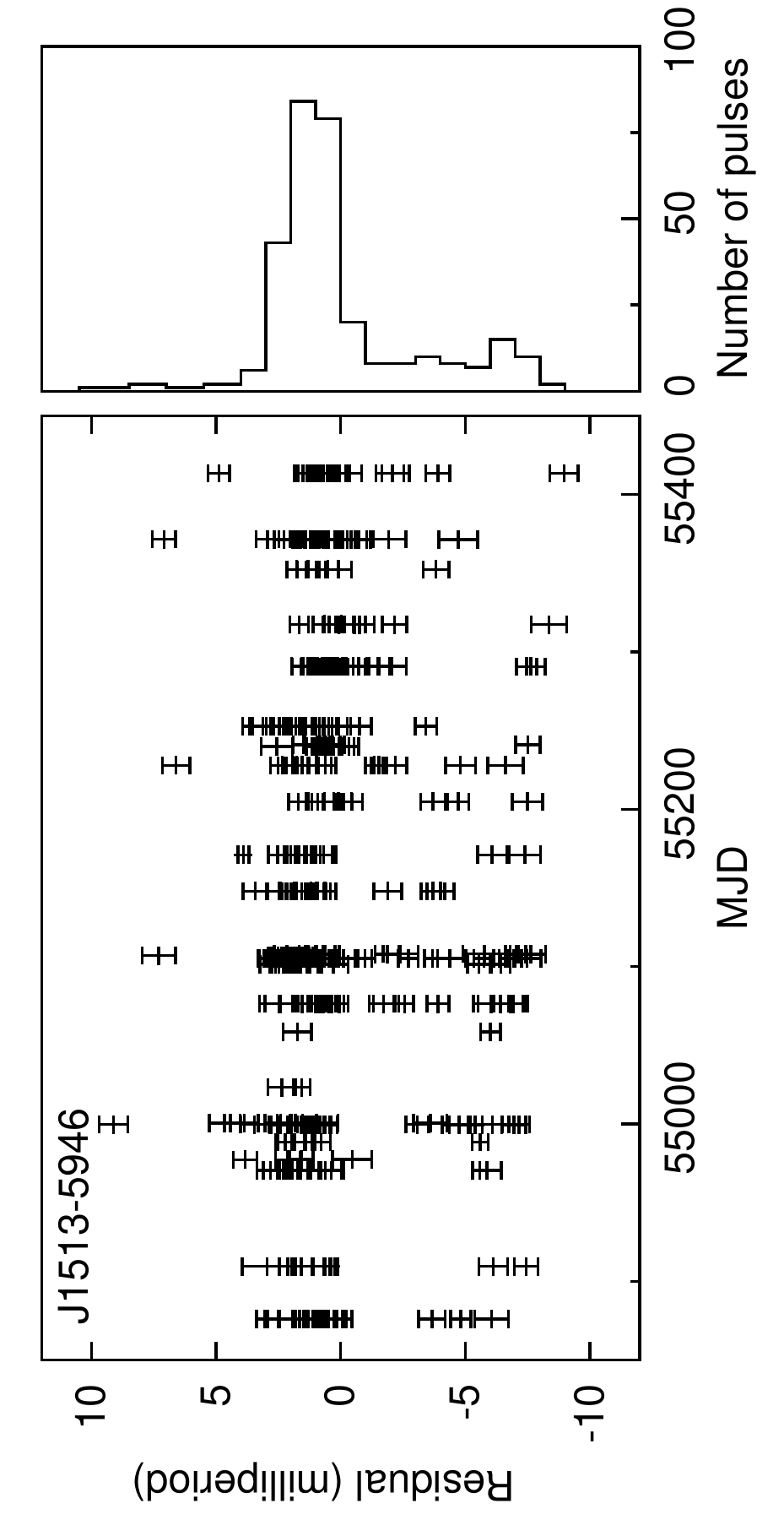}~\includegraphics[trim = 20mm 15mm 165mm 20mm, clip, scale=0.4,angle=-90]{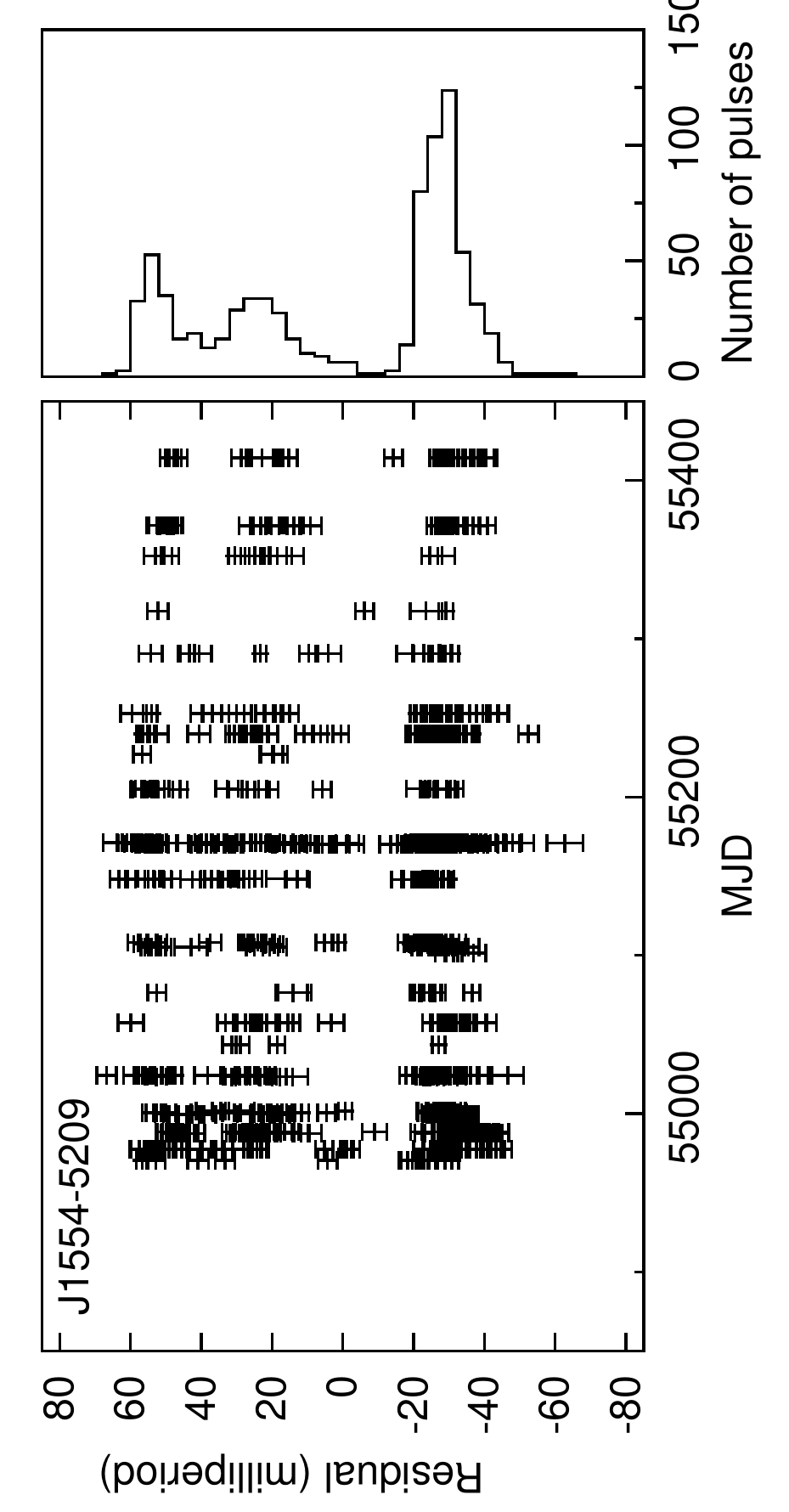}
    \includegraphics[trim = 20mm 15mm 165mm 18mm, clip, scale=0.4,angle=-90]{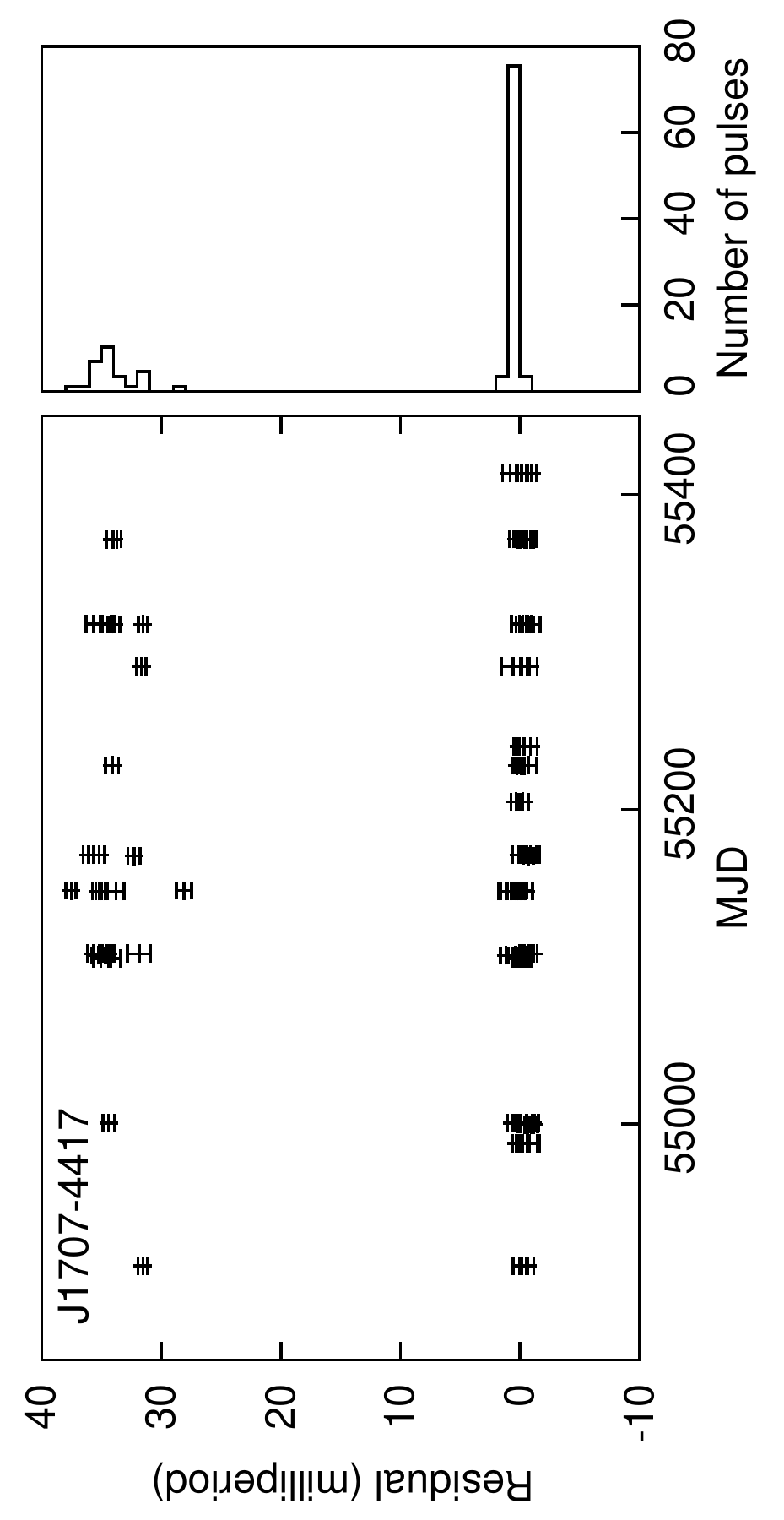}~\includegraphics[trim = 20mm 13mm 165mm 20mm, clip, scale=0.4,angle=-90]{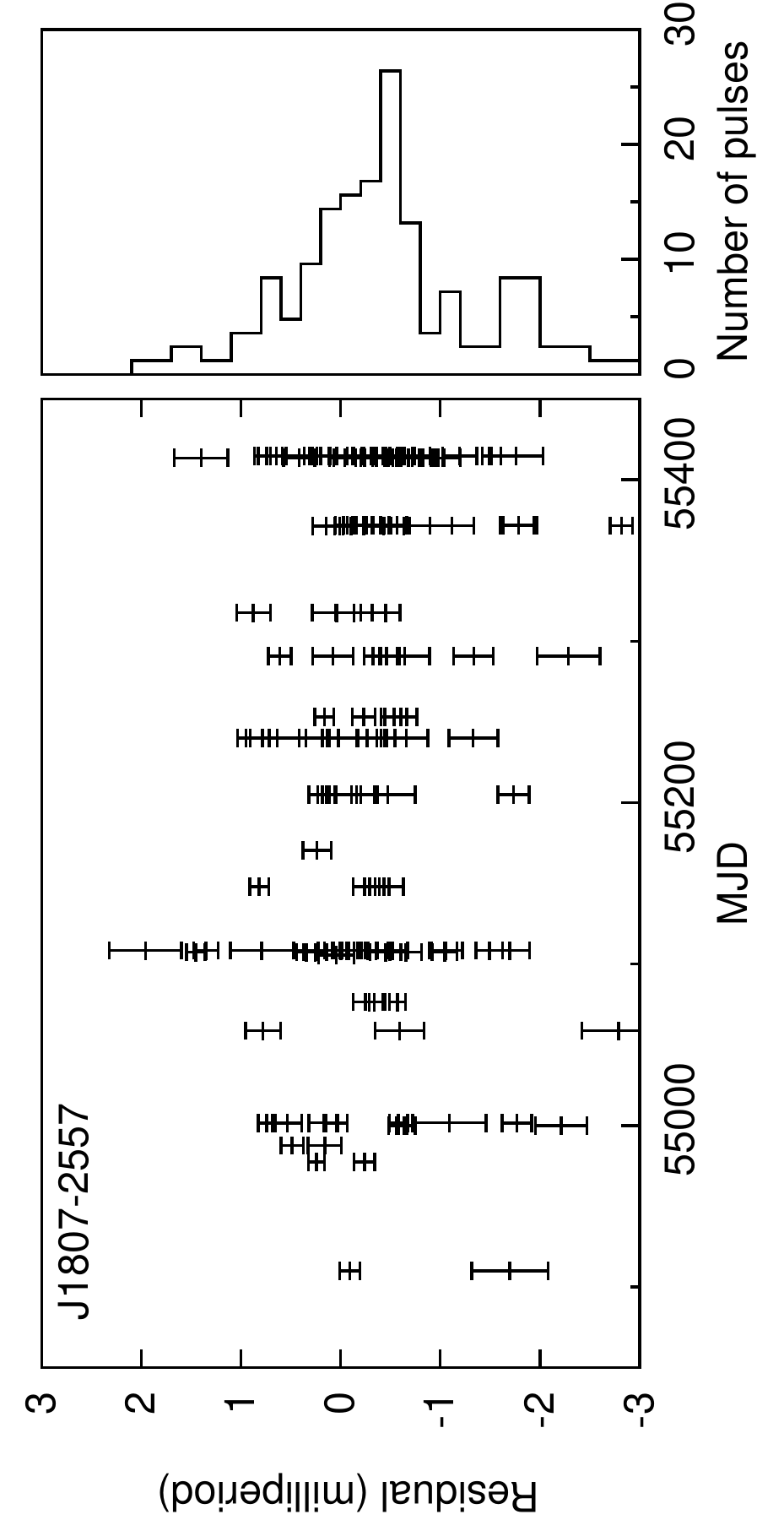}
    \includegraphics[trim = 20mm 15mm 165mm 18mm, clip, scale=0.40,angle=-90]{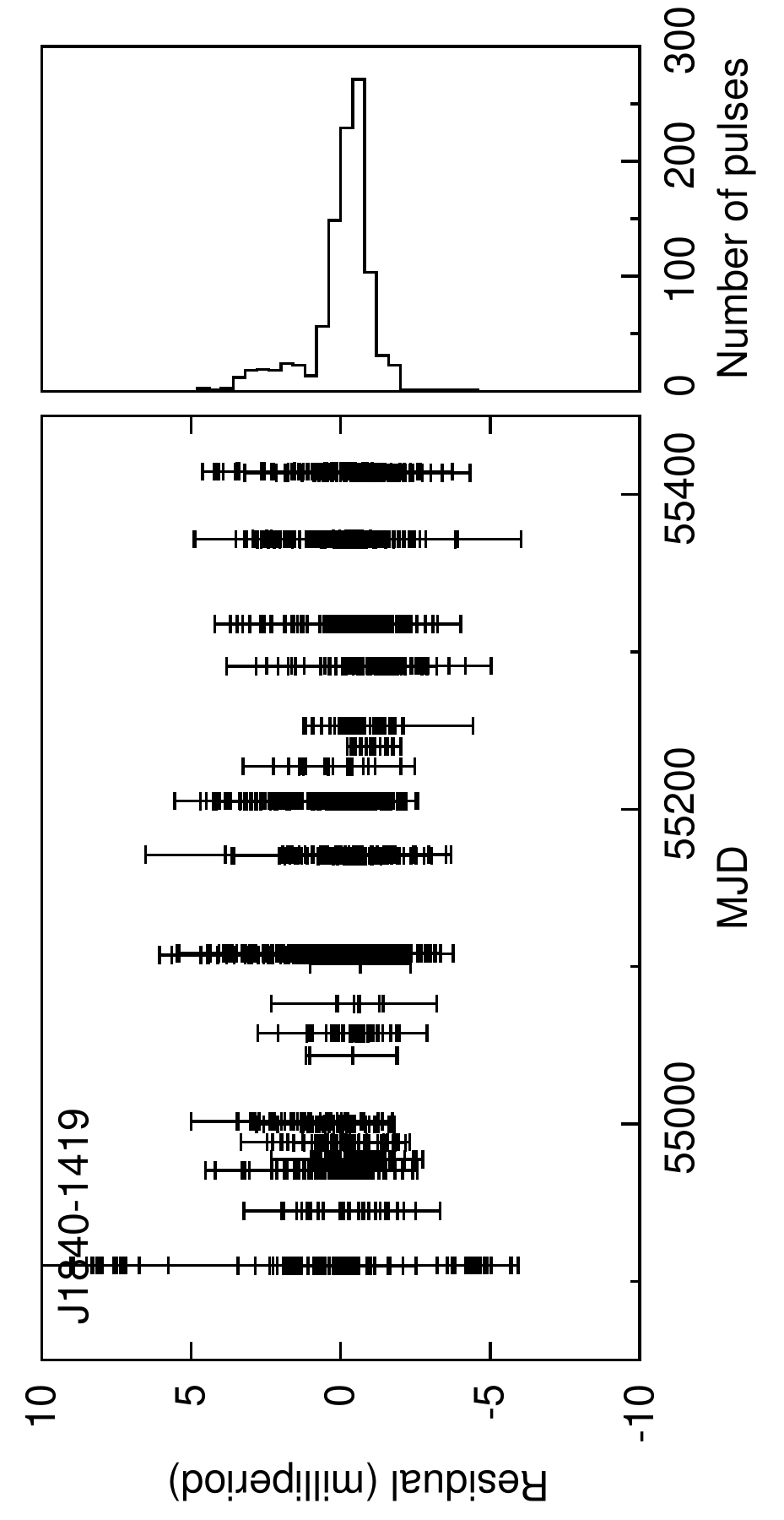}~\includegraphics[trim = 20mm 15mm 165mm 20mm, clip, scale=0.40,angle=-90]{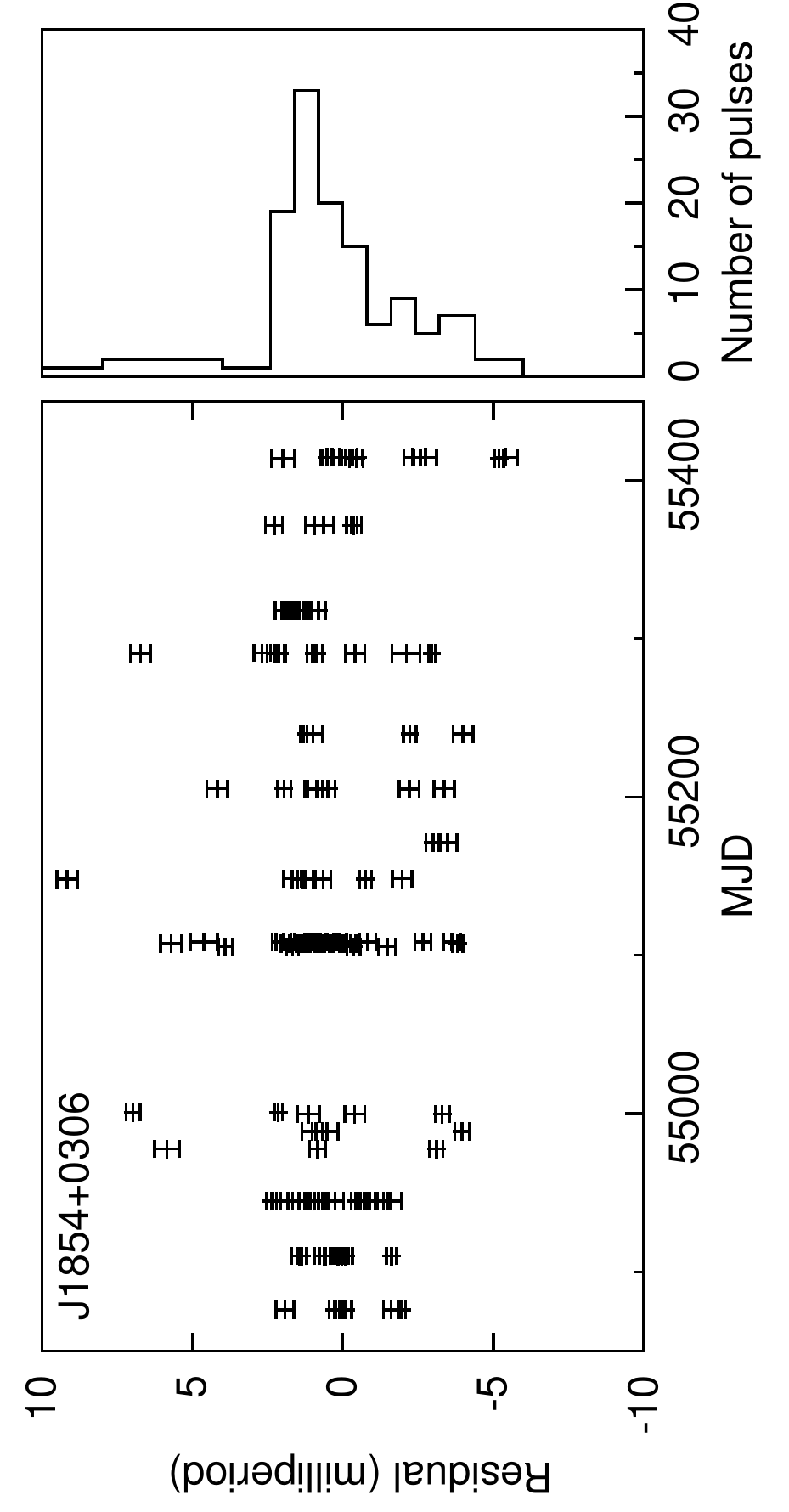}
  \end{center}
  \vspace{-5pt}
  \caption{\small{Plotted are timing residuals for the 6 PMSingle
      sources with determined timing solutions, via single pulse
      timing. From top to bottom the sources are: J1513$-$5946,
      J1554$-$5209, J1707$-$4417, J1807$-$2557, J1840$-$1419, and
      J1854+0306. Note the differing ranges in residual for each
      source. This tells us that, for example, J1840$-$1419 and
      J1707$-$4417 are much better single pulse `timers' than
      J1554$-$5209.}}
  \vspace{-5pt}
  \label{fig:pmsingle_resids}
\end{figure*}

J1513$-$5946 (formerly J1514$-$59) is detected in all 30 observations,
totalling 18~hours. The periodic nulling described in K+10 is detected
in every observation. During the `on' periods, J1513$-$5946 is
detectable in a periodicity search. Figure~\ref{fig:pmsingle_resids}
shows its timing residuals where we can clearly see two bands,
symptomatic of two pulse components. Removing this banding,
i.e. simply applying a jump between the two bands, as done for
J1819$-$1458 in \citet{lmk+09}, we obtain a timing solution with
$\chi^2/n_{\mathrm{free}}=4.2$. The fact that this is not equal to 1
is expected due to our fundamental violation of the stable profile
assumption (see \S~\ref{sec:integrated_profiles}) and is due to the
intrinsic variability of the single pulses, i.e. they are variable in
both phase and pulse width. The `on' times are not long enough, at
$\sim1$~minute, to be able to form stable profiles and result in fewer
TOAs with lower error bars, but with the same scatter as shown in
Figure~\ref{fig:pmsingle_resids}. The timing solution places
J1513$-$59 amongst the `normal' pulsars in the $P-\dot{P}$ diagram
(see Figure~\ref{fig:ppdot_2010}), with perhaps a slightly higher than
average magnetic field strength.

J1554$-$5209 (formerly J1554$-$52) is also detected in all
observations, totalling 13~hours. The timing residuals show three
clear bands, which, upon removal, gives us a timing solution with
$\chi^2/n_{\mathrm{free}}\sim10$, which we again attribute to the
intrinsic variability in the single pulses. In units of pulse periods
it has by far the largest scatter in its residuals. J1554$-$5209 is
also occasionally detectable in periodicity searches, although with
less significance. It has been common (see e.g. \citet{dcm+09}) to
define a quantity $r=(S/N)_{\mathrm{SP}}/(S/N)_{\mathrm{FFT}}$, the
ratio of the single pulse search to FFT search signal-to-noise
ratios. For J1554$-$5209, each observation so far has had $r>1$. It is
noticeable in Figure~\ref{fig:ppdot_2010} as the outlying PMSingle
source with the lowest period, and the highest $\dot{E}$ in our
sample.
It has a typical magnetic field for a pulsar and with $\tau=0.9$~Myr
it is the second `youngest' PMSingle source.

J1652$-$4406 is a radio pulsar with a very large rotation period of
$P=7.707$~s. In fact, J1652$-$4406 is the third slowest radio pulsar
known, just behind J1001$-$5939 ($P=7.73$~s) and J2144$-$6145
($P=8.51$~s). As discussed in \S~\ref{sec:new_discoveries_pmps}, we
have been able to confirm this source since the discoveries announced
in K+10. Although initially identified as a source showing strong
single pulses, we have confirmed it as a periodic source. It is
detected in 22 of 28 followup observations in this way, but never
convincingly re-detected in a search for single
pulses. Figure~\ref{fig:1652_pdmp} shows a typical detection. From
these observations, a timing solution has been obtained with
$\chi^2/n_{\mathrm{free}}=0.75$. The resultant $\dot{P}$ places
J1652$-$4406 just above the death line, just like J1840$-$1419, but
for this source, unlike J1840$-$1419, there is little prospect of high
energy followup as it is towards the Galactic centre, with
$l=341.56\degree$, $b=0.09\degree$ and with
$DM=786\;\mathrm{cm^{-3}\,pc}$ has an inferred distance of
$8.4$~kpc~\citep{cl02}. At $10$ times further distance we expect $100$
times less X-ray flux than from J1840$-$1419, but the situation is
likely to be even worse given the extra absorption that would result
from the large neutral hydrogen density in the Galactic centre.

J1707$-$4417 (formerly J1707$-$44) has been detected in all but one of
23 observations which have totalled 13~hours. The timing residuals
show two clear bands, separated by $\sim200$~ms. There are no other
pulses detected between these two bands although there are instances
where both pulse components are seen together. This suggests that the
active emission time is longer than the time which the emission beam
spends in our line of sight. This is consistent with both the `patchy
beam'~\citep{lm88}, and `hollow cone'~\citep{ran93} beam
models. Removing the banding effect, the timing solution we determine
is remarkably good with $\chi^2/n_{\mathrm{free}}=1.1$, indicating
that the single pulses are very stable in phase and pulse
width. J1707$-$4417 is an old neutron star with $\tau=7.8$~Myr and
lies quite close to the death line, just above J1840$-$1419, in the
$P-\dot{P}$ diagram.

J1807$-$2557 (formerly J1807$-$25) is detected in all observations
covering a total of 16~hours. The timing residuals do not show any
obvious banding, although there is a slight suggestion of a second
band (see Figure~\ref{fig:pmsingle_resids}). The scatter in the
residuals is quite large and the fit has
$\chi^2/n_{\mathrm{free}}\sim20$. Evidently the single pulses from
this source are quite variable in phase. We can also see that the
error bars in the TOAs vary considerably in extent, indicating that
the shape and/or the strength of the individual pulses varies
appreciably between detections. Just as for J1707$-$4417 and
J1840$-$1419, it is an old neutron star with $\tau=8.8$~Myr.

J1840$-$1419 (formerly J1841$-$14) has a large burst rate with strong
single pulses detected at a rate of approximately one per minute. It
can usually be detected in periodicity searches but with less
significance than single pulse searches. Just as for J1707$-$4417, it
has an exceptionally good timing solution with a
$\chi^2/n_{\mathrm{free}}=1.5$, indicating that its single pulses are
very stable in shape and in phase. The proximity of this old pulsar
allows the prospect of X-ray observations. We have recently performed
such observations using \textit{Chandra} and we will report the
results of these observations elsewhere. J1840$-$1419 lies just above
the radio death line and, as such, studies of this star will help us
to investigate important questions concerning old, dying pulsars.

J1854$+$0306 (formerly J1854$+$03) has been detected in 28 of 31
observations during 16~hours of followup. The timing solution has
$\chi^2/n_{\mathrm{free}}\sim40$ and we can see in
Figure~\ref{fig:pmsingle_resids} that the observed scatter is much
larger than the error bars of individual TOAs, indicating variability
in pulse phase. The pulse widths are not seen to vary to the same
degree. Of the PMSingle sources, J1854$+$0306 has the strongest
magnetic field, the second strongest of all the RRAT sources with
determined $B$, behind J1819$-$1458 (see
Table~\ref{tab:pmsingle_timing_solutions}).

For completeness, Table~\ref{tab:pmsingle_timing_solutions} also lists
the 11 PMPS RRATs dicovered in the PMPS and reported in M+06. Since
the discovery of these sources, followup timing observations have been
performed, primarily at Parkes, but also at the GBT, Arecibo and
Jodrell Bank~\citep{mlk+09,lmk+09}. Of these 11, there are seven for
which timing solutions have been obtained are given in the table.

\begin{table*}
  \begin{center}
    \caption{\small{The timing solutions, both complete and
        incomplete, of all the repeating PMPS RRAT sources, as well as
        the BB10 candidates which we have confirmed.  Sources denoted
        with a $\clubsuit$ are those discovered in K+10 and whose
        timing solutions are presented in this work. Sources denoted
        with a $\star$ are those discovered in M+06 and whose timing
        solutions were published in \citet{mlk+09} and
        \citet{lmk+09}. The periods quoted for the BB10 sources are
        those determined from our observations. Columns 2--6 inclusive
        constitute the fitted values. The last column states the DM
        but note that this was not fit in our timing analyses as
        followup observations at several frequencies have not been
        performed. Periods marked with a $\dagger$ have had different
        values published previously. In both cases (J1754$-$30 and
        J1654$-$23) a mis-identification of a terrestrial radio pulse
        as astrophysical is to blame, as has become evident after
        further observations which have revealed many more pulses and
        the true period.}}\label{tab:pmsingle_timing_solutions}
    \begin{tabular}{cccccccc}
      \hline\hline Source & RA & DEC & P & $\dot{P}$ & PEPOCH & Data Span & DM \\
      & (J2000) & (J2000) & (s) & ($10^{-15}$) & (MJD) & (MJD) & ($\rm{cm^{-3}\,pc}$) \\
      \hline

      \multicolumn{3}{l}{\textbf{Complete PMPS Timing Solutions}} \\
      J0847$-$4316$^{\star}$ & 08:47:57.33(5) & $-$43.16:56.8(7) & 5.9774927370(7) & 119.94(2) & 53816 & 52914--54716 & 292.5(0.9) \\
      J1317$-$5759$^{\star}$ & 13:17:46.29(3) & $-$57:59:30.5(3) & 2.64219851320(5)  & 12.560(3) & 53911 & 53104--54717 & 145.3(0.3) \\
      J1444$-$6026$^{\star}$ & 14:44:06.02(7) & $-$60:26:09.4(4) & 4.7585755679(2) & 18.542(8) & 53893 & 53104--54682 & 367.7(1.4) \\
      J1513$-$5946$^{\clubsuit}$ & 15:13:44.78(1) & $-$59:46:31.9(7) & 1.046117156733(8) & 8.5284(4) & 54909 & 54876--55413 & 171.7(0.9) \\
      J1554$-$5209$^{\clubsuit}$ & 15:54:27.15(2) & $-$52:09:38.3(4) & 0.1252295584025(7) & 2.29442(5) & 55039 & 54970--55414 & 130.8(0.3) \\
      J1652$-$4406$^{\clubsuit}$ & 16:52:59.5(2) & $-$44:06:05(4) & 7.707183007(4) & 9.5(2) & 54947 & 54850--55413 & 786(10) \\
      J1707$-$4417$^{\clubsuit}$ & 17:07:41.41(3) & $-$44:17:19(1) & 5.7637770030(4) & 11.65(2) & 54999 & 54909--55371 & 380(10)\\
      J1807$-$2557$^{\clubsuit}$ & 18:07:13.66(1) & $-$25:57:20(5) & 2.76419486975(4) & 4.994(2) & 54984 & 54909--55414 & 385(10) \\
      J1819$-$1458$^{\star}$ & 18:19:34.173(1) & $-$14:58:03.57(1) & 4.26316403291(5) & 575.171(1) & 53351 & 51031--54938 & 196(1) \\
      J1826$-$1419$^{\star}$ & 18:26:42.391(4) & $-$14:19:21.6(3) & 0.770620171033(7) & 8.7841(2) & 54053 & 53195--54909 & 160(1) \\
      J1840$-$1419$^{\clubsuit}$ & 18:40:32.96(1) & $-$14:19:05(1) & 6.5975626227(4) & 6.33(2) & 55074 & 54909--55239 & 19.4(1.4) \\
      J1846$-$0257$^{\star}$ & 18:46:15.49(4) & $-$02:58:36.0(2) & 4.4767225398(1) &  160.587(3) & 53039 & 51298--54780 & 237(7) \\
      J1854$+$0306$^{\clubsuit}$ & 18:54:02.98(3) & $+$03:06:14(1) & 4.5578200962(1) & 145.125(6) & 54944 & 54876--55414 & 192.4(5.2) \\
      J1913+1330$^{\star}$ & 19:13:17.975(8) & $+$13:30:32.8(1) & 0.92339055858(2) & 8.6799(2)  & 53987 & 53035--54938 & 175.64(0.06) \\


      \multicolumn{3}{l}{\textbf{Preliminary/Unsolved PMPS Sources}} \\
      J1047$-$58$^{\clubsuit}$ & 10:47:56(55) & $-$58:41(7) & 1.23129(1) & - & 55779 & - & 69.3(3.3) \\
      J1423$-$56$^{\clubsuit}$ & 14:23:11(53) & $-$56:47(7) & 1.42721(7) & - & 54557 & - & 32.9(1.1) \\
      J1703$-$38$^{\clubsuit}$ & 17:03:26(37) & $-$38:12(7) & 6.443(1) & - & 54999 & - & 375(12) \\
      J1724$-$35$^{\clubsuit}$ & 17:24:43(36) & $-$35:49(7) & 1.42199(2) & - & 54776 & - & 555(10) \\
      J1727$-$29$^{\clubsuit}$ & 17:27:19(33) & $-$29:59(7) & - & - & - & - & 93(10) \\
      J1754$-$30$^{\star}$ & 17:54:16(33) & $-$30:11(7) & 1.32049(1)$\dagger$ & - & 55025 & - & 293(19) \\
      J1839$-$01$^{\star}$ & 18:39:53(29) & $-$01:36(7) & 0.93190(1) & - & 51038 & - & 307(10) \\
      J1848$-$12$^{\star}$ & 18:48:02(30) & $-$12:47(7) & 6.7953(5) & - & 53158 & - & 88(2) \\
      J1911$+$00$^{\star}$ & 19:11:48(29) & $+$00:37(7) & 6.94(1) & - & 52318 & - & 100(3) \\


      \multicolumn{3}{l}{\textbf{Unsolved BB10 RRATs}} \\                                                                                   
      J0735$-$62 & 07:35:24(63) & $-$62:58(7) & 4.865(1) & - & 55352 & - & 19(8) \\
      J1226$-$32 & 12:26:50(34) & $-$32:27(7) & 6.192997(7) & - & 55000 & - & 37(10) \\
      J1654$-$23 & 16:54:03(31) & $-$23:35(7) & 0.54535972(3)$\dagger$ & - & 55261 & - & 74.5(2.5) \\

      \hline
    \end{tabular}
  \end{center}
\end{table*}

\begin{table}
  \begin{center}
    \caption{\small{The derived quantities for the 14 PMPS surces
        discovered as RRATs (the $\clubsuit$ and $\star$ denote
        discovery in K+10 and M+06 respectively), which now have
        coherent timing solutions.  The interpretations of $B$ and
        $\tau$ should be made with caution, as described in
        \S~\ref{sec:discussion}. The values quoted are obtained from
        evaluating
        $B_{\mathrm{vac}}=3.2\times10^{19}\;\mathrm{G}\sqrt{P\dot{P}/\sin^2\alpha}$,
        $B_{\mathrm{ff}}=2.6\times10^{19}\;\mathrm{G}\sqrt{P\dot{P}/(1+\sin^2\alpha)}$
        (with $\alpha=90\degree$ in both cases), $\tau=P/\dot{P}$ and
        $\dot{E}=4\pi^2I\dot{P}P^{-3}$ (see e.g. \citet{ls04,spi06}).
    }}\label{tab:inferred_properties}
    \begin{tabular}{cccc}
      \hline\hline Source & $B_{\mathrm{vac}}$, $B_{\mathrm{ff}}$ & $\tau$ & $\dot{E}$ \\
      & ($10^{12}$~G) & (Myr) & ($10^{31}\;\mathrm{erg\,s^{-1}}$) \\
      \hline

      J0847$-$4316$^{\star}$ & 25.1, 14.1 & 0.8 & 2.0 \\
      J1317$-$5759$^{\star}$ & 6.3, 3.5 & 3.2 & 2.5 \\
      J1444$-$6026$^{\star}$ & 10.0, 5.6 & 4.0 & 0.6 \\
      J1513$-$5946$^{\clubsuit}$ & 3.0, 1.7 & 1.9 & 29.4 \\
      J1554$-$5209$^{\clubsuit}$ & 0.5, 0.3 & 0.9 & 4605.9 \\
      J1652$-$4406$^{\clubsuit}$ & 8.6, 4.8 & 12.8 & 0.1 \\
      J1707$-$4417$^{\clubsuit}$ & 8.3, 4.7 & 7.8 & 0.2 \\
      J1807$-$2557$^{\clubsuit}$ & 3.8, 2.1 & 8.8 & 0.9 \\
      J1819$-$1458$^{\star}$ & 50.1, 28.2 & 0.1 & 32.8 \\
      J1826$-$1419$^{\star}$ & 2.5, 1.4 & 1.3 & 79.4 \\
      J1840$-$1419$^{\clubsuit}$ & 6.5, 3.7 & 16.5 & 0.1 \\
      J1846$-$0257$^{\star}$ & 25.1, 14.1 & 0.4 & 6.3 \\
      J1854$+$0306$^{\clubsuit}$ & 26.0, 14.6 & 0.50 & 6.1 \\
      J1913$+$1330$^{\star}$ & 2.5, 1.4 & 1.6 & 39.8 \\


      \hline
    \end{tabular}
  \end{center}
\end{table}

\subsection{Preliminary/Unsolved Sources}\label{sec:preliminary_timing}
In addition to the 14 PMPS sources now with coherent timing solutions
there are nine others which have been re-detected on multiple
occasions but do not yet have a coherent timing solution. We review
the status of these in turn.

J1047$-$58 is a very sporadic source, whose rate of detected bursts
varies between extremes an order of magnitude higher and lower than
its average rate of $\sim 4\;\mathrm{hr}^{-1}$. This results in it
being detected in only a quarter of observations, which has hampered
efforts to determine a coherent timing solution, although a solution
is expected for this source, with sufficient observation time.

J1423$-$56 is more stable in its burst rate than J1047$-$58, and
although a solution is not yet determined, it is expected that this
will be possible in the coming months.

J1703$-$38 is another source with a low burst rate of
$\lesssim2\;\mathrm{hr}^{-1}$. Despite this, since its discovery in
K+10, we have been able to determine a period of $P=6.443$~seconds
using period differencing. A timing solution has not been forthcoming
however as very long observations ($> 1$~hr) are needed to guarantee
the detection of multiple pulses (essential for identifying the
topocentric period in each observation). Higher sensitivity
observations and perhaps lower frequencies (where the burst rate might
be higher, as seen by \citet{mac09}) are planned for the future.

J1724$-$35 was the first PMSingle candidate to be confirmed. It has
been missed in 6 of 21 followup observations which have totalled
15~hours. Furthermore, when detected its observed burst rate is
$\lesssim3\;\mathrm{hr}^{-1}$, which is quite low, so that obtaining a
coherent timing solution has not been possible. A renewed attempt will
be made in the future to `solve' this source, using higher
sensitivity, again at lower frequencies, and this is planned for
future work.

J1727$-$29 has by far the lowest burst rate of any of our confirmed
sources with just 4 pulses detected in 6 hours. Further followup is
not feasible given the required telescope time, as such a low rate
makes determining a timing solution very difficult. In fact we have
not even determined the underlying period, if any, in this
source. With pulses of $\sim7$~ms wide its maximum source size is
constrained to be $\sim2100$~km by causality. This is much larger than
a neutron star but less than the minimum radius for a relativistic
white dwarf at the Chandrasekhar mass~\citep{st83} so that we suspect
a neutron star origin.

Of the other PMPS RRATs, we are confident that timing solutions will
be obtained for J1754$-$30, J1839$-$01 and J1848$-$12, but it a
solution for J1911$+$00 seems unlikely due to its very low burst rate.
Of the three BB10 RRATs which we have confirmed it seems that timing
solutions should be possible with continuing observations.

J0735$-$62 is not detected in two followup observations, each of ten
minutes duration. Recently we have made a third, 30-minute observation
where it was easily detected, and thus confirmed it for the first
time, with 20 strong single pulses. Analysing the TOA differences we
determine a topocentric period of $P=4.865(1)$~s, consistent with the
initial estimate of $P=4.862$~s period published in BB10. Additionally
the two non-detections support their claim that the source suffers
from severe scintillation. For this reason we do not yet know if
obtaining a timing solution for this source is possible using a
reasonable amount of observing time, but a single lengthy observation
is planned in the coming months, to investigate this very question.

In the original observation of J1226$-$32 only 3 pulses were
detected, but this was sufficient for BB10 to derive a period of
$P=6.193$~s. We have confirmed this candidate and have observed 45
pulses in almost 3~hours of followup, although in one third of the
observations it is not detectable. We confirm the published period,
and our provisional timing solution is coherent since January 2010
and regular ongoing observations should reveal a full timing solution
for this source.

The original detection of J1654$-$23 also consisted of just 3
pulses. We have confirmed this source and have determined a period of
$0.545$~s, which differs from the published estimate of BB10. This is
not very surprising given their small number of detected
pulses. Interestingly, the period we determine, from 106 pulses
detected in 2.3~hours, is not at a different harmonic. This suggests
that perhaps one of the 3 pulses initially identified was terrestrial
in origin. As for J1226$-$32 we have a provisional timing solution,
coherent since January 2010 and regular observations are ongoing.

In addition to the above three sources, we have attempted to confirm
three other sources. We have observed J0923$-$31 and J1610$-$17 for
1.0 and 1.2 hours respectively but have not been able to make a
confirmation. We have detected 5 weak pulses from J1753$-$12 at the
correct DM, during 1.3~hours of observation, although we hope a more
significant confirmation will come with time. We have not yet followed
up these 3 sources for as long as the 3 new confirmations. This is, in
some sense, by design, as these sources showed just 1, 1 and 3 pulses
respectively in their discovery observations, so we decided to
initially focus on the higher burst rate source (which were
subsequently confirmed).

\subsection{PMPS Timing Status}
Of the 30 sources now identfied in the PMPS, 23 have been re-detected
on multiple occasions, of which 22 of these have known periods and 14
of these now have coherent timing solutions. Of the unsolved sources,
the prospects for obtaining solutions are promising for five of these,
but seem unlikely for the rest, due, primarily, to the low rate of
pulse detection, i.e. an unfeasibly long observing time would be
required. Additionally, two of the BB10 RRATs are expected to have
full solutions in due course. We also note that these sources are the
only radio neutron stars with timing solutions obtained using
individual pulses, rather than averaged
profiles. Figure~\ref{fig:ppdot_2010} shows an up to date $P-\dot{P}$
diagram showing all known radio pulsars, the 14 RRATs, the magnetars
and the XDINSs, for which $P$ and $\dot{P}$ are known.

\begin{figure*}
  \begin{center}
    \includegraphics[trim = 20mm 20mm 0mm 20mm, clip, scale=0.3,angle=0]{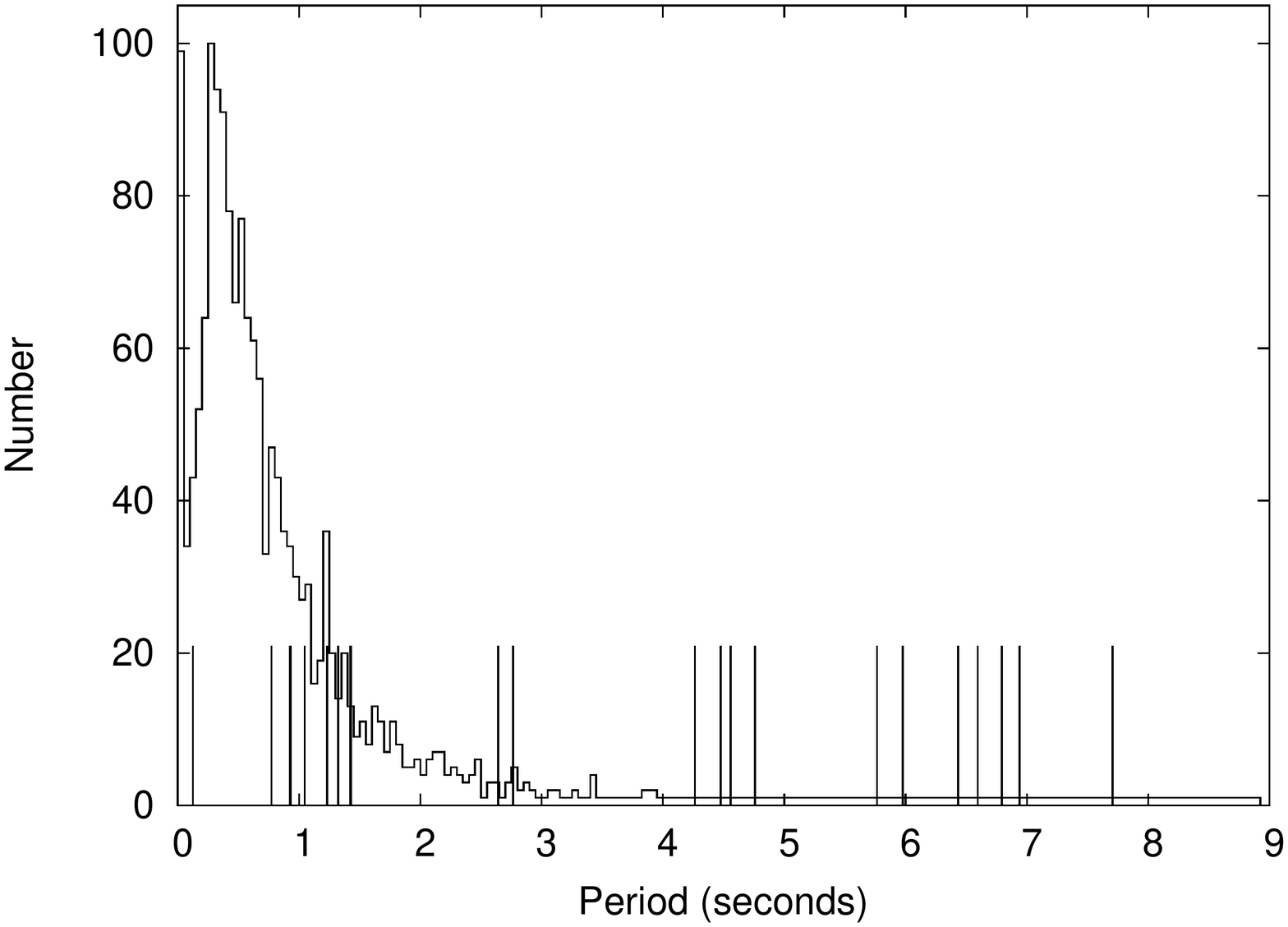}~\includegraphics[trim = 20mm 20mm 0mm 20mm, clip, scale=0.3,angle=0]{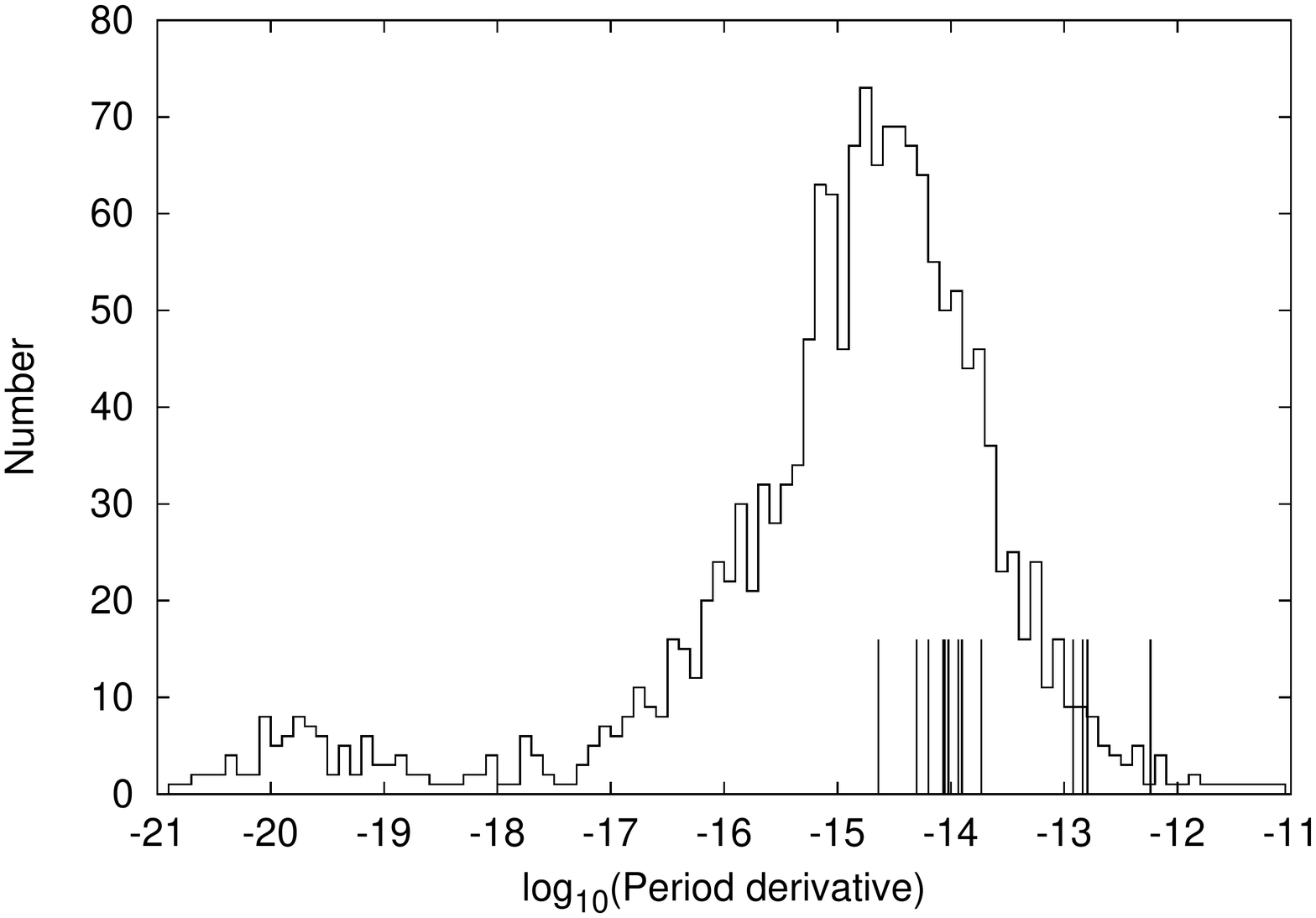}
    \includegraphics[trim = 20mm 20mm 0mm 20mm, clip, scale=0.3,angle=0]{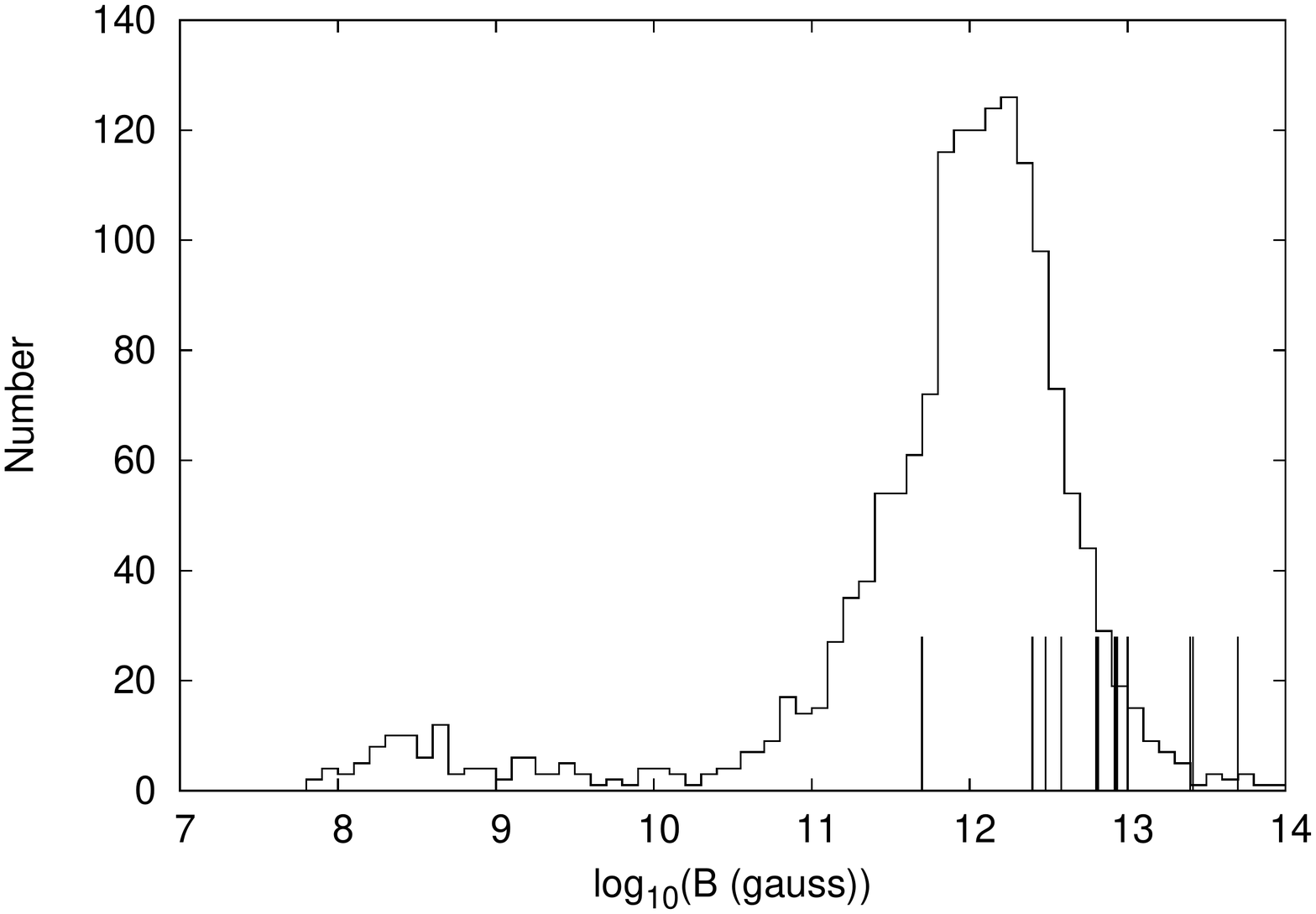}~\includegraphics[trim = 20mm 20mm 0mm 20mm, clip, scale=0.3,angle=0]{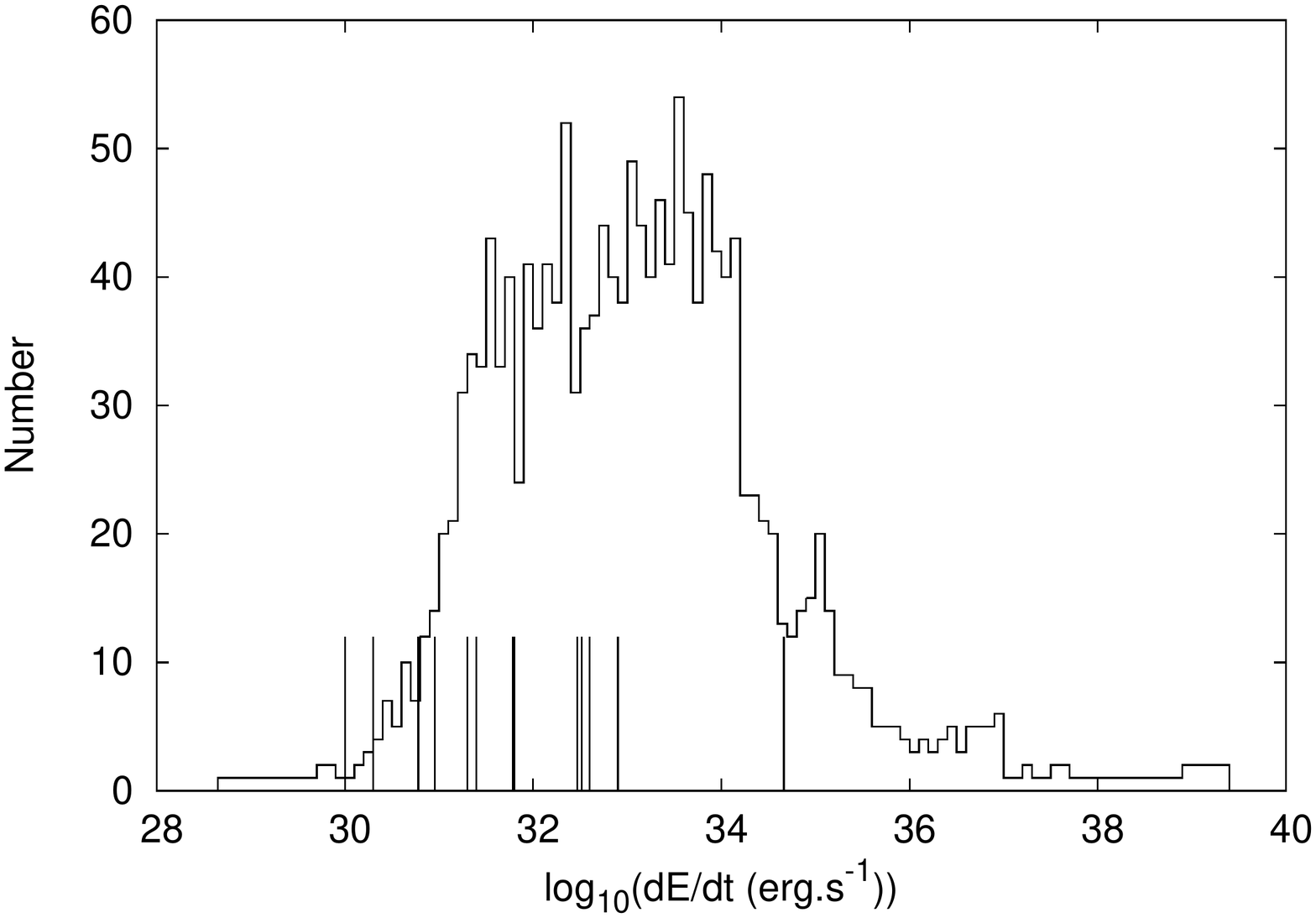}
    \includegraphics[trim = 20mm 20mm 0mm 20mm, clip, scale=0.3,angle=0]{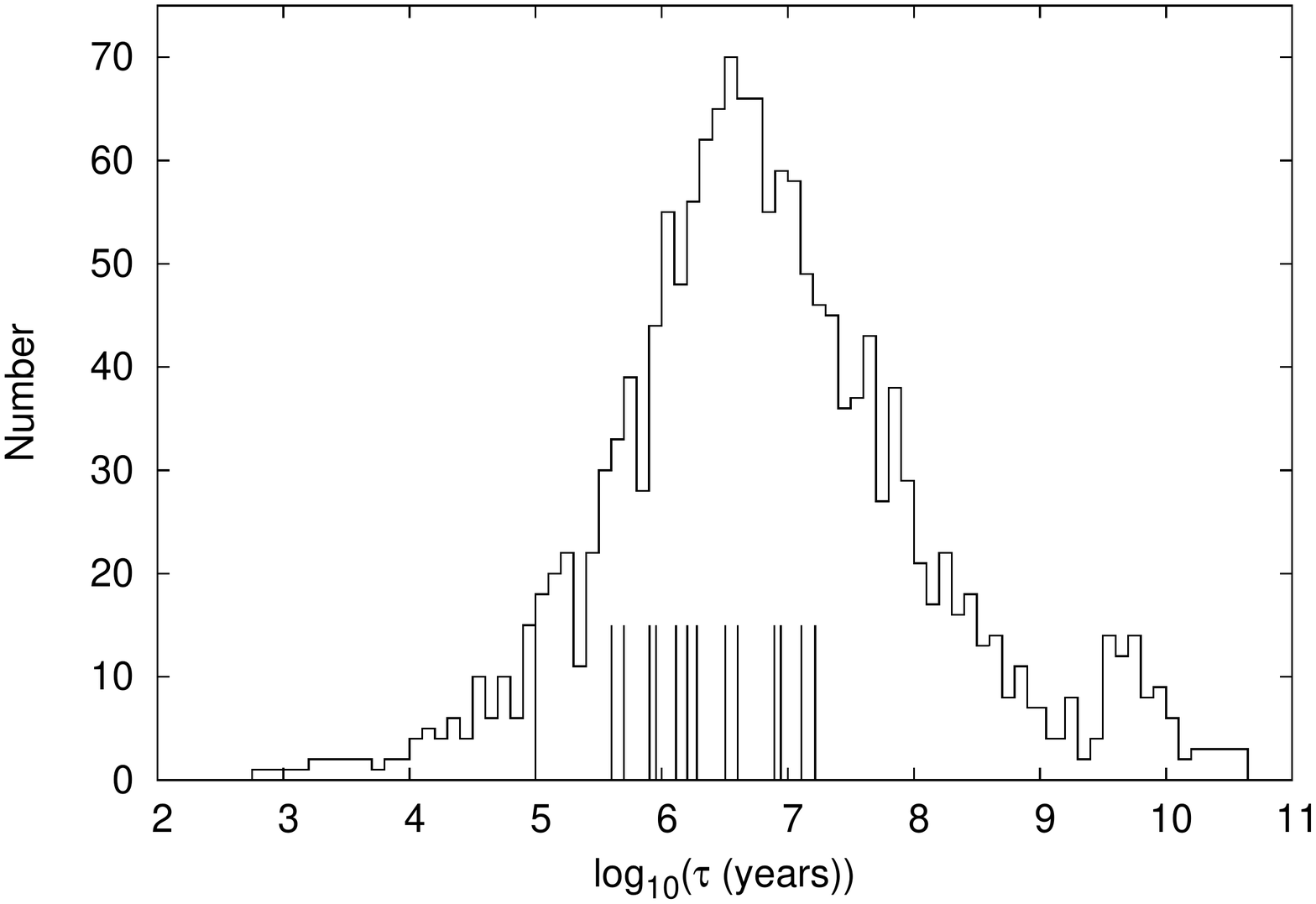}~\includegraphics[trim = 20mm 20mm 0mm 20mm, clip, scale=0.3,angle=0]{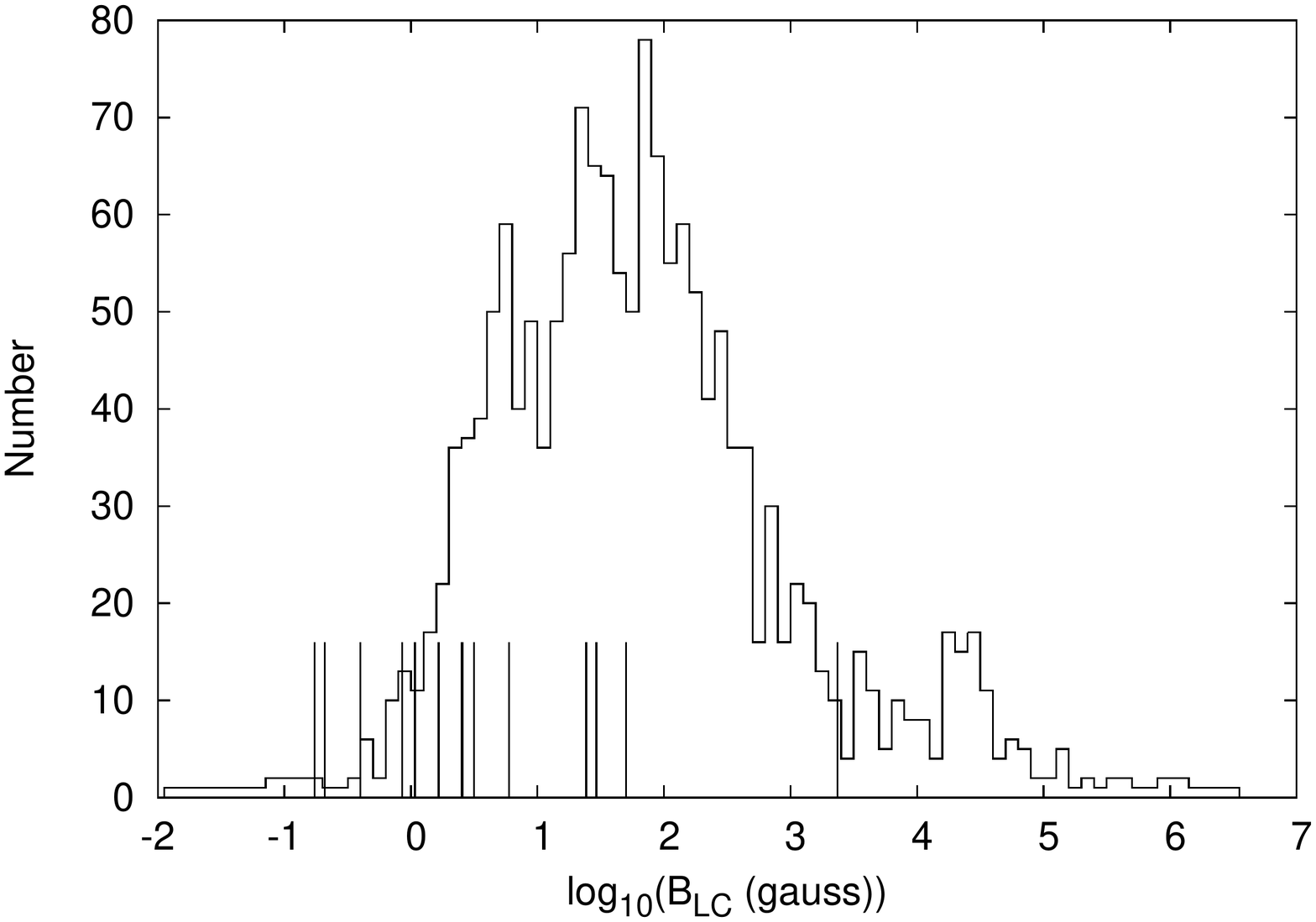}
    \includegraphics[trim = 20mm 20mm 0mm 20mm, clip, scale=0.3,angle=0]{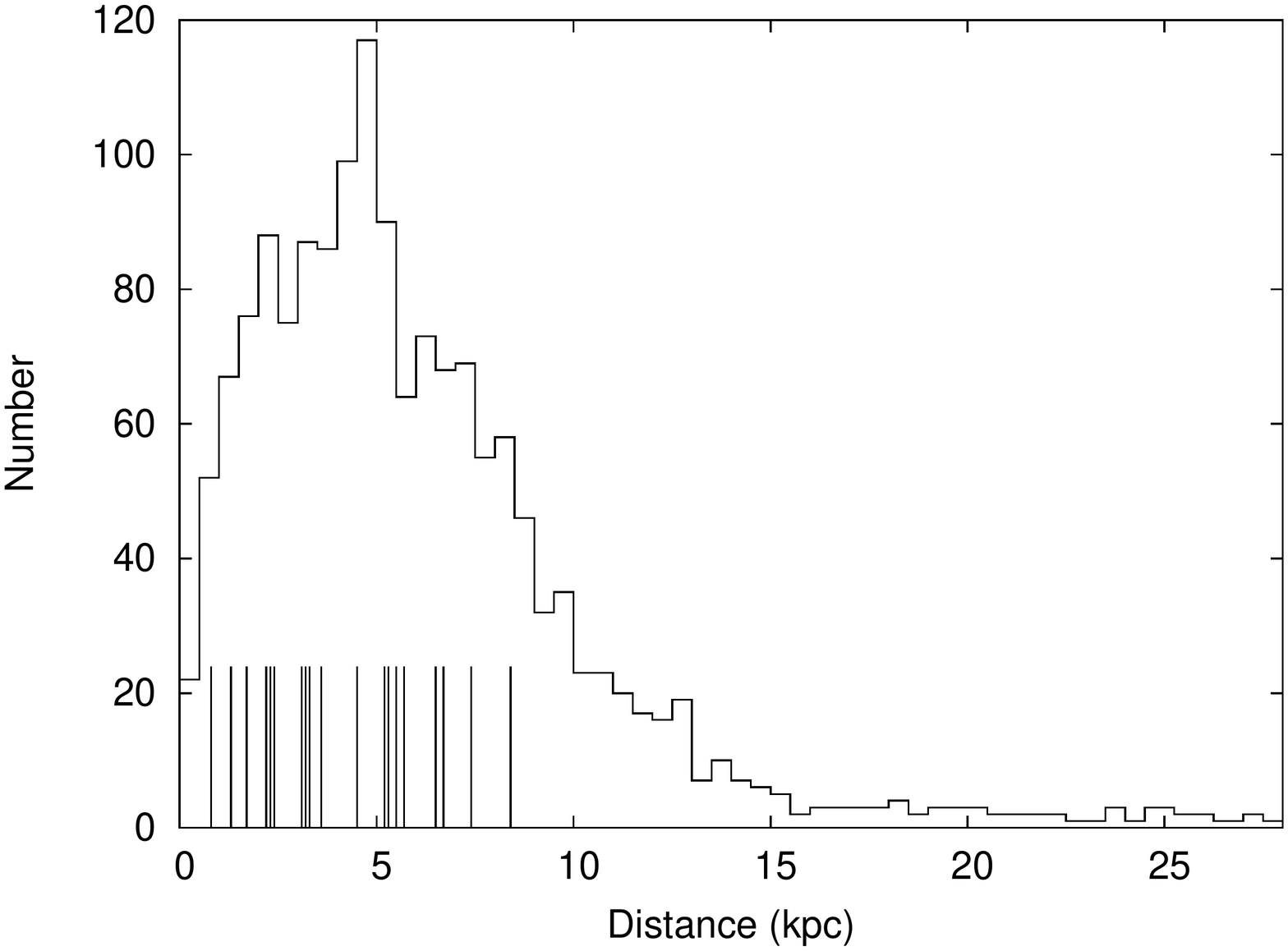}~\includegraphics[trim = 20mm 20mm 0mm 20mm, clip, scale=0.3,angle=0]{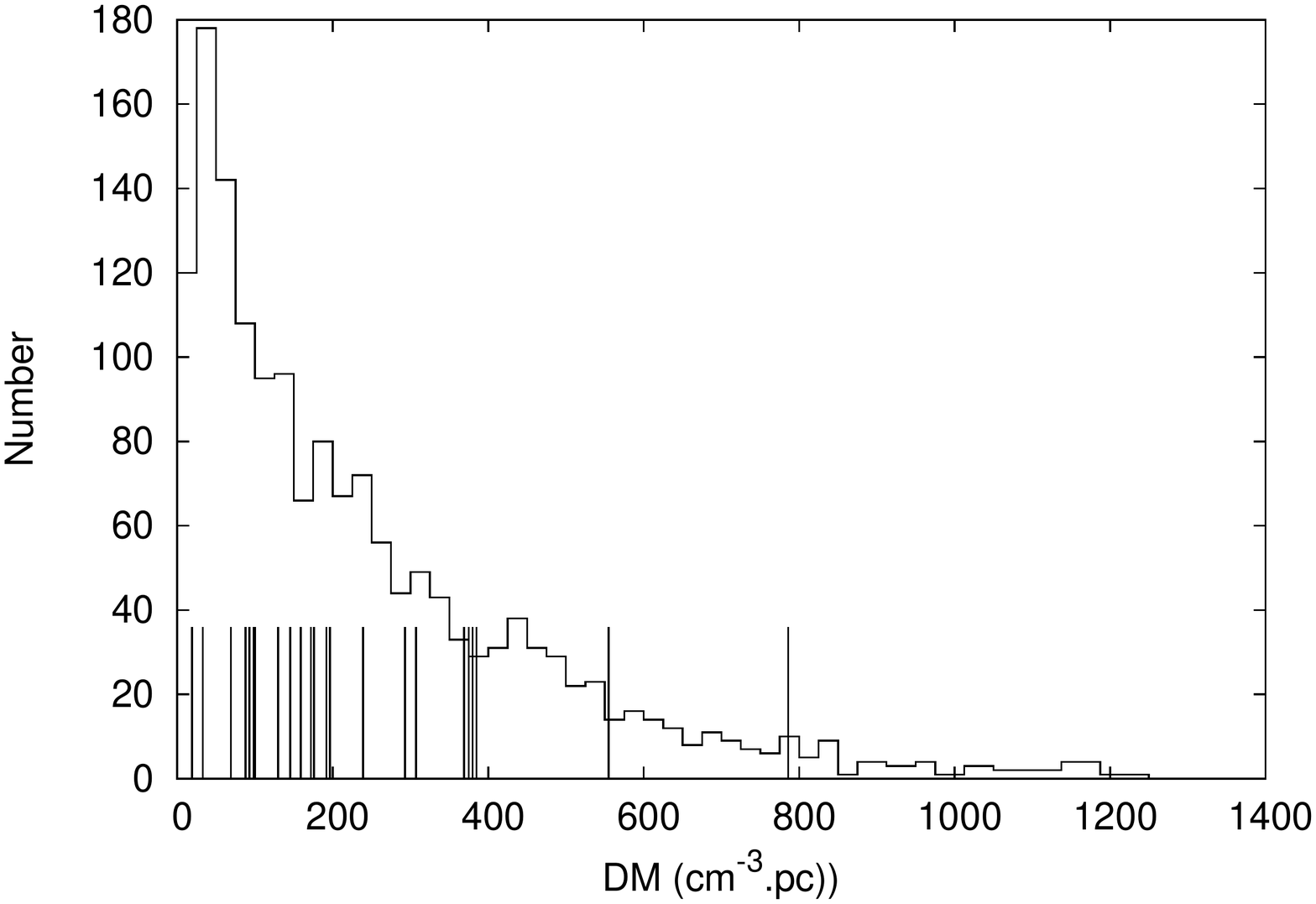}
  \end{center}
  \vspace{-10pt}
  \caption{\small{In each panel we compare the properties of the known
      pulsar population to those of the RRATs. The parameters are $P$,
      $\dot{P}$, $B$, $\dot{E}$, $\tau$, $B_{\mathrm{LC}}$, distance
      and DM. Apart from $P$, distance and DM the abscissa is plotted
      as a base-10 logarithm.}}
  \label{fig:all_pmps_rrat_properties}
\end{figure*}

\begin{figure*}
  \begin{center}
    \includegraphics[trim = 20mm 20mm 10mm 30mm, clip,scale=0.55,angle=0]{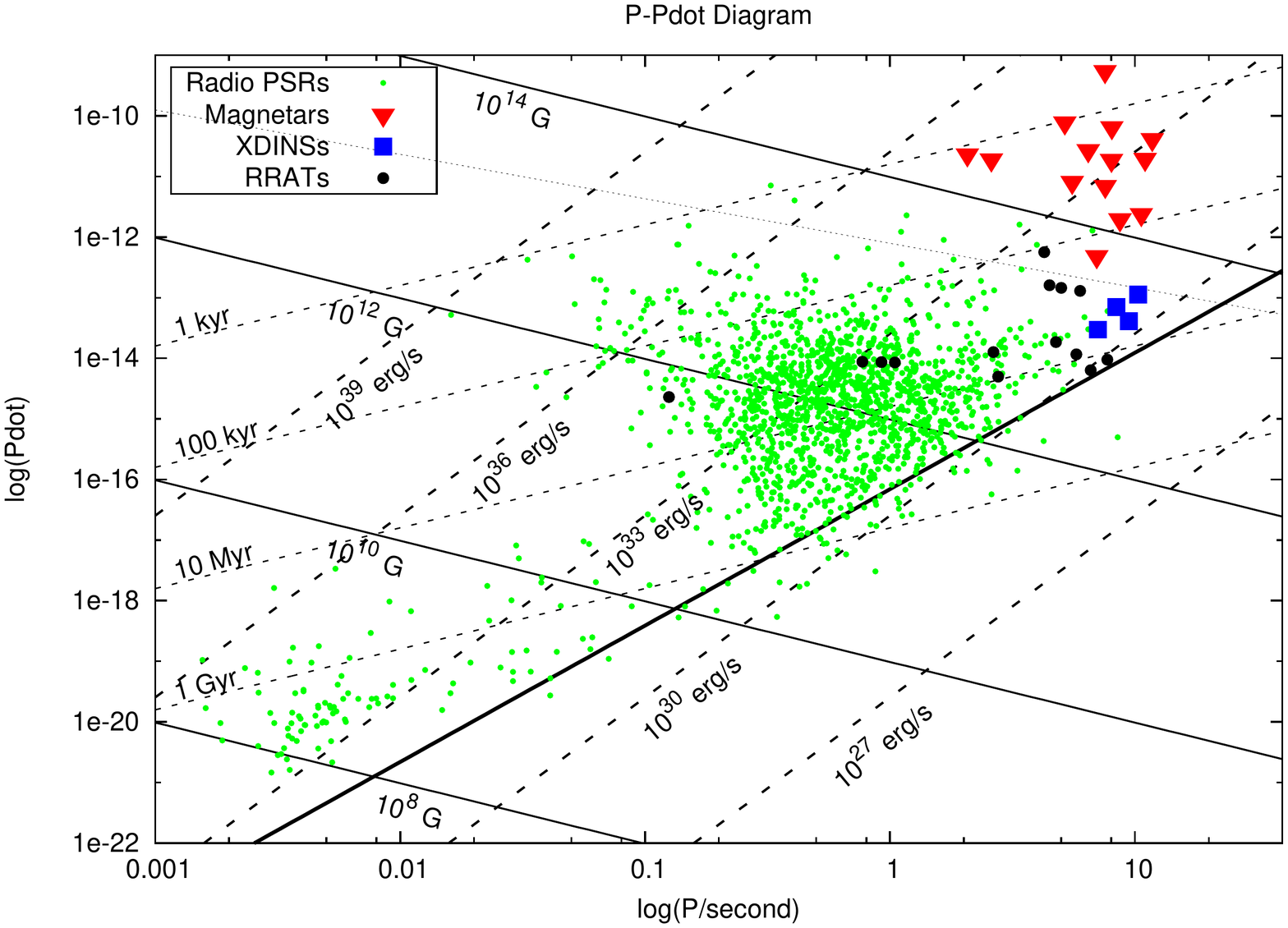}
  \end{center}
  \vspace{-5pt}
  \caption{\small{The pulsar $P-\dot{P}$ diagram. Shown are the radio
      pulsars, which can clearly be seen to consist of two classes ---
      the `slow' pulsars and the MSPs, as well as those RRATs, XDINSs
      and magnetars with known period derivative. The region in the
      bottom right (bounded by the thick black line) denotes the
      canonical `death valley' of \citet{cr93a} where we can see there
      is a distinct (but not complete) lack of sources. The radio
      loud-radio quiet boundary of \citet{bh98}, which divides the
      magnetars and the XDINSs, is also shown (dotted line) and we can
      see that only $\sim1\%$ of sources are found above this
      line. Also plotted are lines of constant $\dot{E}$, $B$ and
      $\tau$, calculated using the standard equations~\citep{ls04}.}}
  \vspace{-5pt}
  \label{fig:ppdot_2010}
\end{figure*}

Figure~\ref{fig:all_pmps_rrat_properties} summarises the properties,
both measured and derived, resulting from the timing analysis. The
values for all PMPS RRATs are shown and contrasted with the
distribution of values in the pulsar population as a whole. We can see
from Figures~\ref{fig:all_pmps_rrat_properties} and
\ref{fig:ppdot_2010} that the RRATs certainly have long periods with
half of the the 22 sources having periods $P>4$~s. The four sources
with the highest inferred magnetic field strengths occupy a void
region of $P-\dot{P}$ space and J1819$-$1458 remains the source with
the highest $B$. At least four other RRATs (and possibly six) of the
14 with known $\dot{P}$ are `normal'. The remaining four sources have
very long periods and lie just above the death line. Identification of
these 3 `groups' hints at an answer (or rather answers) to the
question: what is a RRAT? We discuss this in detail in
\S~\ref{sec:discussion} but it is clear that some are normal pulsars,
some old/dying pulsars and some occupy the high-B void region of
$P-\dot{P}$ space.

\section{Discussion}\label{sec:discussion}

In our PMSingle analysis we have discovered 19 sources, which brings
the number of PMPS RRATs known to 30. These discoveries are broadly
consistent with the initial population estimate for RRATs --- we
removed the effects of `RFI blindness'~\citep{ekl09}, which affected
$\sim 1/2$ of the PMPS pointings, and (more than) doubled the known
PMPS RRATs. In addition to the PMPS, others have identified sources to
be RRATs --- seven at Arecibo, including J1854$+$0306~\citep{dcm+09},
two at GBT~\citep{hrk+08,blm+10}, 25 others at Parkes (BB10, as
mentioned already, as well as the first results of the HTRU
surver~\citet{bjj+11}), one at Puschino~\citep{skdl09} and four at
Westerbork~\citep{rub10}. In total, this amounts to 67 sources
identified as RRATs, at the time of writing. Thus, we can see that the
birthrate problem~\citep{kk08} remains and RRATs must be explained
within the context of known neutron star classes. Fortunately, this is
possible.

\subsection{When a Pulsar is a ``RRAT''}
We define a RRAT as:

\textbf{Definition:} \textit{A RRAT is a repeating radio source, with
  underlying periodicity, which is more significantly detectable via
  its single pulses than in periodicity searches.}

This (arbitrary) definition is clearly a detection-based definition
and a source can only be labelled a RRAT for a specific
survey/telescope/observing frequency/observing time\footnote{In fact
  the RRAT label is not permanent: a source may be detected as a RRAT
  but subsequently be more easily detected in periodicity
  observations, even for identical observing setups. This was the case
  for the PMSingle source J1652$-$4406.}. It says nothing directly
about the intrinsic properties of the source --- we feel that this is
appropriate. Thus: \textit{an observing setup might be contrived so as
  to make any pulsar a RRAT.} Are the group of RRATs, so defined, in
any way special? In a general sense, where any observational setup is
possible, they are not, but for realistic survey specifications, they
can be. Single-pulse searches make a selection on the parameter space
of possible sources. The group of RRATs resulting from this may be of
interest, for a number of reasons, as we will elucidate.

\subsection{Selection Effects}
\begin{figure*}
  \vspace{-10pt}
  \begin{center}
    \includegraphics[trim = 0mm 0mm 0mm 0mm, clip,scale=0.45,angle=0]{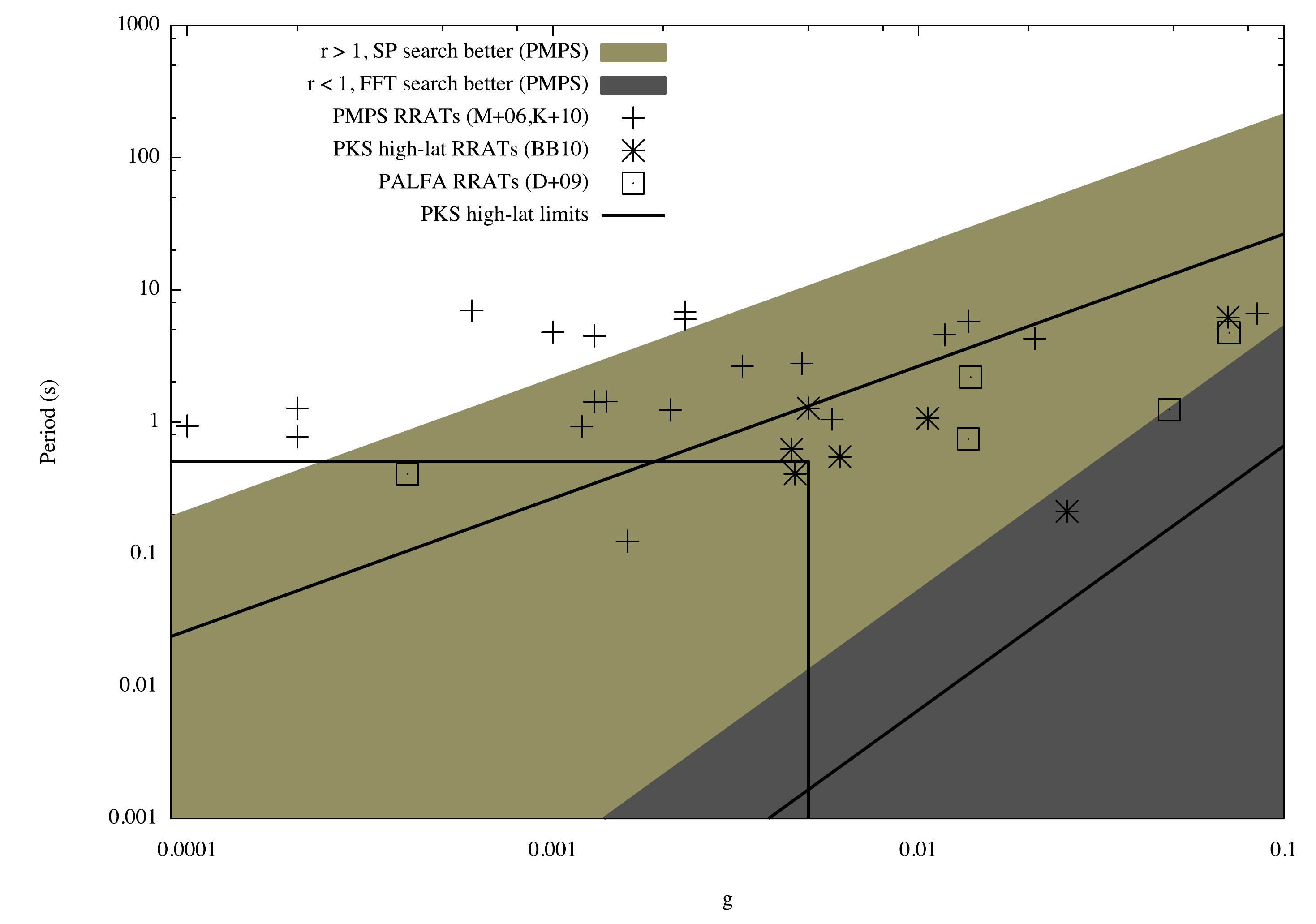}
  \end{center}
  \vspace{-10pt}
  \caption{\small{Plotted is $g-P$ space, where $g$ is the fraction of
      periods where a pulses is detected and $P$ is the rotation
      period. Considering the ratio
      $r=(S/N)_{\mathrm{SP}}/(S/N)_{\mathrm{FFT}}$ the regions where
      SP searches (light grey+white, $r>1$) and FFT searches (dark
      grey, $r<1$) are more effective for the PMPS are marked,
      defining ``RRAT'' and ``pulsar'' regions. Over-plotted are the
      PMPS RRATs with measured periods as reported in M+06 \&
      K+10. Also plotted are the boundaries (black lines) for the
      sources reported by BB10 with known $P$ and $g$. We also plot
      the sources reported in \citep{dcm+09} (D+09 in the
      figure). J1854$+$0306 is plotted with the PMPS sources, although
      it was also identified in PALFA. We note that the boundaries for
      the inner-Galaxy PALFA pointings are the same as for the Parkes
      high-latitude surveys if we assume no difference in
      sensitivity. This is of course incorrect, and due to this extra
      difference (the Parkes surveys have the same sensitivity as each
      other) the D+09 sources are plotted simply for illustration.}}
  \vspace{-10pt}
  \label{fig:gP}
\end{figure*}

We begin by considering what this definition means as far as selection
effects are concerned. As an example, we can take a source, with
period $P$, which emits (detectable) pulses a fraction of the time $g$
and nulls (or is not detected) a fraction of the time $1-g$. Then we
can use the well-known selection effect in $g-P$ space for this
scenario~\citep{mc03,kea10a}, namely $r>1$ when $Tg^2/4<P<Tg$, where
$T$ is the observing time. For a given $g$, the low period limit
defines the $r=(S/N)_{\mathrm{SP}}/(S/N)_{\mathrm{FFT}}=1$ condition,
so that, at lower periods an FFT search is more effective. For higher
periods than $Tg$ there is unlikely to be even one pulse during the
observation. Figure~\ref{fig:gP} shows a plot of $g-P$ space with
`RRAT-PSR' boundaries marked for the 35-minute pointings of the
PMPS. Here we are using our definition of `RRAT', and using
`pulsar/PSR' as a synonym for `more easily, or only detectable in a
periodicity search'. Thus PMPS RRATs are those sources in the light
grey or white shaded regions. Different surveys will have different
RRAT-PSR boundaries, e.g. the higher-latitude Parkes surveys analysed
by BB10 had shorter pointings and hence different boundaries which are
over-plotted on Figure~\ref{fig:gP}. Thus the `RRAT' J1647$-$36
detected in the high-latitude surveys would have been detected as a
`pulsar' if it were surveyed in the PMPS. We note that, in reality,
the $g$ values we measure represent the \textit{apparent} nulling
fraction, i.e. the intrinsic values of $g$ may be higher depending on
the pulse-to-pulse modulation and distance to the source~(Weltevrede
et al. 2006; BB10)\nocite{wsrw06,bb10}. Periodicity searches also make
a selection in $g-P$ space, the dark grey region of
Figure~\ref{fig:gP}. In comparison to periodicity searches,
single-pulse searches are sensitive to high period sources ($\gtrsim
10$~s) with moderate nulling fraction ($\sim 0.1$) down to very short
period ($\sim10^{-3}-10^{-1}$~s) sources with large nulling fraction
($10^{-4}-10^{-3}$).

From inspecting Figure~\ref{fig:gP}, we can make a number of
remarks. Firstly, we can see that the average `RRAT' and `PSR' periods
we infer would be:
\begin{align}
  \langle P \rangle_{\mathrm{RRAT}}&=\frac{\int P(\int
  RRAT(g,P)dg)dP}{\int\int RRAT(g,P)dgdP} \;, \\
  \langle P \rangle_{\mathrm{PSR}}&=\frac{\int P(\int
  PSR(g,P)dg)dP}{\int\int PSR(g,P)dgdP} \; ,
\end{align}
where $RRAT(g,P)$ and $PSR(g,P)$ are distribution functions in $g-P$
space. For a uniform $g-P$ distribution these simply correspond to the
shaded areas in the figure, and the results can be easily
calculated. For sensible ranges (the ranges plotted in
Figure~\ref{fig:gP}, for $P<10$~s, say) we always get $\langle P
\rangle_{\mathrm{RRAT}} > \langle P \rangle_{\mathrm{PSR}}$. It would
not then be useful (or fair) to compare period distributions of
sources selected in these ways. Further examining the figure we can
see that the bottom left-hand corner (bounded by the black lines in
the figure) is lacking in sources. Moving upwards a decade in $P$ for
the same $g$ range (say) we expect to get $\sim10$ times as many
sources, if the distribution is uniform, and this is, roughly, what we
see. Going up another decade in $P$ we do not see a further increase
in sources, most likely due to there being no radio-visible pulsars
with $P\gtrsim 10$~s. The period distribution is approximately uniform
in $\log P$ in the band $\sim0.5-8$~s (given the small numbers of
sources),
which we contrast with the lognormal distribution for pulsars centred
at $0.3$~s~\citep{rl10a}.

The distribution in $g$ may be of more interest. We can see that,
within the band where we see sources, the distribution is not uniform,
but looks somewhat uniform in $\log g$. We can thus explain the
distribution of sources as follows: (i) The low $P$--low $g$ region is
devoid of sources as this does not represent a large area of parameter
space and/or there are not many sources with these characteristics;
(ii) The $P\gtrsim10$~s region does not have any active radio pulsars,
consistent with what is expected for slow pulsars which have passed
the death line; (iii) The $P\sim0.5-8$~s region for
$g\sim10^{-4}-10^{-1}$ shows a somewhat uniform distribution in $\log
g$, suggesting that there are more RRAT-selected pulsars with high
nulling fractions than would be expected from a uniform distribution
in $g-P$ space. To turn this around, if we search for RRATs, we are
likely to find pulsars with high nulling fraction. These data are not
sufficient to identify any trend in $g$ with $P$, and there are less
data for investigating any relationships with $\dot{P}$, $\tau$, $B$,
$\dot{E}$, etc. As a final comment on Figure~\ref{fig:gP} we note that
there are several PMPS sources just above the light grey region. Here
it is unlikely that there will be a pulse during a 35-minute pointing
but nevertheless there are 8 sources. For each of these, which we were
lucky to detect, we might expect there are several similar sources,
which we missed, simply due to probability. This is yet another
argument, if any were needed, in favour of surveying the sky multiple
times. Indeed the HTRU survey~\citep{kjs+10} is currently surveying
the Galaxy at declinations $\delta < 10\degree$, which includes the
region covered by the PMPS, and has already identified new sources
which were presumably `off' during the 35-minute PMPS
pointings~\citep{bbj+11}.

Slow-down rate, $\dot{P}$, is not subject to any selection effect in
either RRAT or periodicity searches, as typical $\dot{P}$ values have
no effect during survey pointings. Looking at
Figure~\ref{fig:all_pmps_rrat_properties} we can see that the
$\dot{P}$ values for 5 RRATs in particular are higher than average,
with high corresponding magnetic field strengths of $B\gtrsim
10^{13}$~G (less than $4\%$ of the overall population have $B$ values
this high). Excluding J1554$-$5209 the other sources all have slightly
higher than average magnetic fields with $B>10^{12}$~G, consistent
with the earlier claim of \citet{mlk+09}.  As single-pulse searches
have no selection effects against high nulling fraction pulsars, and
these same sources seem to have high-$B$ values, this suggests the
question: Do long period and/or high-$B$ pulsars have higher nulling
fraction? Here we reach a dead end as the nulling properties of
pulsars are completely unknown in the $B\sim B_{\mathrm{QC}}$ and
$P\gtrsim 3$~s regions, where a number of RRATs are found. One reason
for this is that these regions have a dearth of sources and in fact
the PMPS RRATs represent a significant fraction of the known sources
in these regions. As the PMPS RRATs are not obviously very distant we
also ask the question: Do long period and/or high $B$ pulsars have
large modulation indices?  \citet{wes06,wse07} suggest a weak
correlation of modulation index with $B$, but again, the number of
high-$B$ and long period sources in this sample was small.

Another selection effect that the PMPS RRATs suffered from is the
`low-DM blindness' of the original single-pulse search, i.e. the
possibility that low-DM sources were missed due to the effects of
RFI. 
Our re-analysis removed this effect and in fact discovered a number of
low-, as well as high-DM sources which had initially been missed due
to RFI (see e.g. Figure 1 of K+10) so that we believe this selection
effect has been largely removed.

\subsection{Explanations of sporadic behaviour}
There have been many `solutions' proposed as to how a sporadic
emission mechanism might operate, sometimes involving trigger
mechanisms for (re-)activating pulsar emission due to transient
disturbances in fall-back discs~\citep{li06}, surrounding asteroidal
material~\citep{cs08} or plasma trapped in radiation
belts~\citep{lm07}. However, as we have asserted that RRATs are merely
pulsars which fit a particular selection criteria, for a given
observational setup, the question of a solution becomes more a
question of what types of pulsars are we most likely to detect as
RRATs. There are two obvious types consistent with high observed
nulling fractions: (i) weak/distant pulsars with high modulation
indices; (ii) nulling pulsars. BB10 have dubbed ``objects which emit
only non-sequential single bursts with no otherwise detectable
emission at the rotation period'', as `classic RRATs', but by this
definition, there may be no RRATs (see \S~\ref{sec:misconceptions}) so
we do not use this terminology.

The projected population of RRATs is not as high if some sources are
covered by solution (i). Such sources will have low-luminosity
periodic emission. The pulsar population is estimated only above some
threshold luminosity, typically $L_{\mathrm{min}}\sim
0.1\;\mathrm{mJy\,kpc}^2$, so that if these sources are above
$L_{\mathrm{min}}$ they are already accounted for within
low-luminosity selection-effect scaling factors in estimates of the
pulsar population~\citep{lfl+06,rl10a}. If the underlying periodic
emission were below $L_{\mathrm{min}}$ then these sources would
contribute to a birthrate problem by increasing the pulsar population
estimate. In fact the required low-luminosity turn-over\footnote{There
  must be a low-luminosity turn-over so that the integral $\int
  N(L)dL$ does not diverge at the low end. Here $N(L)dL$ denotes the
  number of pulsars with luminosity between $L$ and $L+dL$.} is not
yet seen, which is why artificial cut-offs are usually applied in
population syntheses (see e.g. \citet{fk06}). BB10 argue that extreme
modulation can account for all but two RRATs, but notably not
J1819$-$1458 and J1317$-$5759, which agrees with recent analysis by
Miller et al. (2011, submitted). The true number covered by scenario
(i) may be smaller as it assumes analogues of the extreme source
PSR~B0656+14 to be common in the Galaxy (BB10). So it appears that a
number of RRATs
are accounted for by scenario (i), whereas some are not, and seem to
fit type (ii).

\subsection{Switching Magnetospheres?}\label{sec:switching_magnetospheres}
Scenario (ii), which sees RRATs as nulling pulsars, extends the
boundaries of observed nulling behaviour. In comparison to the
previously observed nulling sample, RRATs would be considered extreme,
with nulls of minutes to hours, as opposed to $\sim$seconds. Excluding
the RRATs, nulling has been observed in $\sim50$ pulsars, but, if we
include pulsars where an upper limit on the nulling fraction has been
obtained, the number in the literature is
$\sim100$~\citep{big92a,viv95a,lcx02,fsk+04,rwr05,wes06,wmj07}. Of
these, there are $50$ with $P>1$~s, 10 with $P>2$~s and 1 with
$P>3$~s. The nulling behaviour of long period and high-B sources is
completely unknown. Some authors have claimed a correlation of nulling
fraction with period~\citep{big92a}, whereas others have claimed the
correlation is instead with characteristic age~\citep{wmj07}. Some of
the observed RRATs are high-B sources with long period, but are young
in terms of characteristic age. Others are `dying' pulsars having both
long periods and old characteristic ages. Observations of a large
sample of pulsars, selected as RRATs, could then be ideal for the
purpose of testing these competing claims.

Thus, we have `nulling pulsars', with nulls of $1-10$ periods, `RRATs'
with nulls of $10-10^4$ periods and `intermittent pulsars' with nulls
of $10^4-10^7$ periods. It appears that there may be a continuum of
null durations in the pulsar population. The question of the `RRAT
emission mechanism' is then subsumed by the questions of what makes
pulsars null, and why such a wide range of null durations are
possible. Another question of immediate interest is in what cases do
nulls occur --- high-B, long period, old pulsars? Also unexplained are
the non-random~\citep{rr09} and periodic behaviour seen in several
sources, e.g. 1-minute periodicity for PSR~J1819$+$1305~\citep{rw08},
several minutes for the PMSingle source J1513$-$5946, hours for
PSR~B0826$-$34~\citep{dll+79}, $\sim1$~day for PSR~B0823$+$26
(N. Young, private communication) and $\sim1$~month for
PSR~1931$+$24~\citep{klo+06}. 
Considering the more general case of moding, we can add the pulsars
reported by \citet{lhk+10}, which switch between (at least) 2 modes,
with associated switches in spin-down rate. PSR~J0941$-$39 is observed
to switch between `RRAT-like' and `pulsar-like' modes (BB10). Recently
PSR~J1119$-$6127 has been observed to switch between RRAT, pulsar and
null states~\citep{wje10}. Interestingly these changes, in the case of
PSR~J1119$-$6127, are seen to occur contemporaneously with the
occurence of an `anomolous glitch', i.e. one resulting in a net
decrease in spin-down rate, as seen only in RRAT J1819$-$1458
previously~\citep{lmk+09}.

The mounting evidence suggests that it is a general property of (at
least some) pulsars, that they can switch back and forth between two
stable states of emission. 
We note that, as $\dot{E}_{\mathrm{radio}}\ll \dot{E}$, the simple
switching on or off of the radio emission should not result in any
noticeable effect\footnote{Consider a simple calculation for a pulsar
  with radio flux density of $10$~mJy, a distance $1$~kpc away. Its
  radio pseudo-luminosity is then
  $10^{-2}\;\mathrm{Jy\,kpc}^2\approx10^{11}\;\mathrm{W\,Hz}^{-1}$. Assuming
  a constant flux density over a GHz bandwidth gives a luminosity of
  $E_{\mathrm{radio}}=10^{20}\;\mathrm{J.s^{-1}}=10^{27}\;\mathrm{erg\,s^{-1}}$
  which we can compare to the much higher $\dot{E}$ values reported in
  Table~\ref{tab:pmsingle_rrat_properties}.} in $\dot{\nu}$. The fact
that $\dot{\nu}$ changes have been observed in very-long duration
nullers (Kramer et al. 2006\nocite{klo+06}, the effect is unobservable
in short-duration nullers) suggests a large-scale change in the
magnetosphere, i.e. the nulls are not due to the micro-physics of the
emission mechanism~\citep{tim09}. Within the framework of force-free
magnetospheres, it has been shown that a number of stable solutions
are possible with different sizes of the closed field line
region~\citep{con05,tim06}. These solutions are derived as for the the
original solution of \citet{ckf99}, but without the assumption that
the angular velocity of the field lines is equal to that of the
star. \citet{tim09} has shown how moderate changes in the beam shape
and/or current density can cause large changes in $\dot{E}$, and hence
$\dot{\nu}$. For a pulsar changing between two stable states, the
observed emission along our line of sight will change, and this will
be seen as a mode switch. A null will result if the beam moves out of
our line of sight as a result of the switch, or, if there is a
sufficient change in current such that the emission
ceases~\citep{tim09}. \citet{con05} have shown that a sudden depletion
of charges will result in such a change of state (which they refer to
as a `coughing magnetosphere'), but with no explanation for why this
depletion might arise. A recent suggestion by \citet{rmt11} is that
the required change in charge density might be triggered by non-radial
oscillations of the stellar surface, although no driving mechanism for
such oscillations is yet known. What is clear from the data is that
pulsars can switch between stable states. Such an effect, if truly a
generic property of pulsars, can explain the phenomena of moding,
nulling and RRATs. The theoretical work shows that different stable
magnetospheric states exist. The reason why a pulsar would switch
between two states (in particular with a periodicity) is unknown.




\section{Conclusion}\label{sec:conclusion}

\subsection{Facts about RRATs}\label{sec:misconceptions}
We now address a number of assertions, claims and misconceptions
concerning the characteristics of RRATs, that we have encountered
during the last few years. Firstly, the assumption that all RRATs have
high magnetic field strengths is incorrect. If we arbitrarily define
high-B as $B\geq 10^{13}$~G, then there are 5 (or 4) RRATs in this
category using minimum $B$ values for the vacuum (or force-free)
case. Neither is it true that RRATs and magnetars are linked in some
way, despite the tentative link suggested for J1819$-$1458 due to its
unusual glitches. Although true of the data accumulated up to the
original discovery paper of M+06, it is no longer a true statement to
say that RRATs are only detectable in single-pulse searches. Several
of the original, PMSingle and BB10 sources are detectable in
periodicity searches, in some cases occasionally and in some cases
reliably. Similarly the pulse arrival times do not seem to be
random. We have discussed non-random behaviour (here, and in K+10) in
the PMSingle sources, where clustering of pulses is seen, e.g. in
J1724$-$35 and J1513$-$5946. This is also seen in J1913$+$1330 at
Jodrell Bank and at Parkes~\citep{mac09,kea10a}. Consecutive pulses
from RRATs are seen quite often. We detect consecutive pulses in
several PMSingle sources, and in our ongoing observations of
J1819$-$1458 and J1913$+$1330 at Jodrell Bank, and Palliyaguru et
al. (2011, submitted) report higher instances of doublets, triplets
and quadruplets than would be expected by random chance. Palliyaguru
et al. also report an instance of detecting pulses from J1819$-$1458
for 9 consecutive periods. This drastically changes the `activation
timescales' needed in some models (although not all, see
e.g. \citet{zgk07}) of RRAT emission, from $\sim3$~ms to
$\sim35$~s. We can also say that none of the RRATs discovered, which
have coherent timing solutions, are in binary systems.

\subsection{Questions \& Future Work}
Our studies of RRATs raise a number of questions and suggest a number
of lines of enquiry for future work. For instance, we do not know the
significance of the anomalous glitches seen in both J1819$-$1458 and
J1119$-$6127, both pulsars lie in the $B\sim B_{\mathrm{QC}}$ region
of $P-\dot{P}$ space, although there is evidence linking these
glitches with changes in radio emission behaviour~\citep{wje10}. It
would seem that an investigation of all sources (and indeed searches
for more) in this region is warranted. With the discoveries of neutron
stars which switch between 2 or more stable states, it is timely, and
necessary, to perform a complete census of nulling pulsars across the
$P-\dot{P}$ diagram, as nulling properties are known for only a
relatively small fraction of the pulsar population, and are unknown
for high-B and long-period sources. The cause of nulling is unknown:
does slowing down below a critical rotation rate, or a magnetic field
growing/decaying to a certain value, signal the onset of nulling?


It is unknown what decides whether neutron stars with similar spin
properties will manifest themselves as RRATs, as opposed to (say) a
magnetar or an XDINS. The answer to this important question is
fundamental if there is to be ``grand unification of neutron
stars''~\citep{kas10}, i.e. the determination of some kind of
evolutionary framework. For example, the region of $P-\dot{P}$ space
defined by $P=4-10$~s, $\dot{P}=10^{-13}-10^{-12}$ contains radio
pulsars (some `normal' pulsars, some RRATs like J1819$-$1458),
magnetars and XDINSs. For very similar spin-down properties we have
very different observational manifestations. We might speculate that
these different classes, although having similar properties now, have
evolved in completely different ways and may have completely different
ages. The snapshot we see now where these sources seem similar, may
then be mis-representative of their overall evolutionary
behaviour. Alternatively, the conditions for coherent radio emission
may be very sensitive, with this region a particular area of parameter
space on the threshold for emission. This is perhaps consistent with
the transient radio emission seen in magnetars and the extreme nulling
of the RRATs in this region. If the re-connection rate at the Y-point
(separating open and closed field lines in the magnetosphere, see
e.g. \citet{spi06}) were slow, or progressed in steps, then bursts of
radio emission may be expected between dormant phases, when the
magnetospheric configuration was favourable.
A natural explanation for periodic switching between stable
magnetospheric states is still lacking.

The many transient searches underway highlight more basic questions
also, such as what can we use as a reliable estimate of age for
neutron stars (important if an evolutionary framework is ever to
exist), how many neutron stars are there in the Galaxy, and how many
sources remain to be discovered in the archives of existing pulsar
surveys as yet undiscovered. What we do know is that those pulsars
discovered as RRATs are now beginning to represent a significant
number, yet there does not seem to be real cause for concern regarding
the expected number of such sources being discovered. The emission
seen in RRATs does not seem remarkable, other than in its sporadicity,
so that there is no need to formulate any new emission
mechanisms. They can be explained within the existing pulsar
framework, or rather, the existing framework of open
questions. Interestingly, with single-pulse searches, we have a means
with which to identify pulsars which have been difficult to find, in
particular the high-B and the dying pulsars.

\section*{Acknowledgments}
EK acknowledges the FSM, the support of a Marie-Curie EST Fellowship
with the FP6 Network ``ESTRELA'' under contract number
MEST-CT-2005-19669, STFC RG R108020 and helpful discussions about
pulsar timing with Mark Purver \& Kuo Liu. MAM is an Alfred P. Sloan
Fellow and is supported by the Research Corporation. 
The Parkes radio telescope is part of the Australia Telescope National
Facility which is funded by the Commonwealth of Australia for
operation as a National Facility managed by CSIRO.

\end{document}